\documentclass[review,12pt]{elsarticle} 

\usepackage[colorlinks, citecolor =blue]{hyperref}
\hypersetup{linkcolor=blue}

\usepackage[english]{babel}
\usepackage{amsmath}
\usepackage{graphicx} 
\usepackage{color}
\usepackage{longtable}
\usepackage{multirow}
\usepackage{hyperref}
\usepackage{booktabs}
\usepackage{morefloats}
\usepackage{textcomp}
\usepackage{gensymb}
\usepackage{epstopdf}
\usepackage{geometry}
\usepackage{caption}
\usepackage{subfig} 
\usepackage{amsmath}
\usepackage{textcomp}
\usepackage{amsfonts}
\usepackage{mathtools}
\usepackage{booktabs} 
\usepackage{array}
\usepackage[table,xcdraw]{xcolor}
\usepackage{colortbl}
\usepackage{float}
\usepackage{makecell}
\usepackage{color}
\usepackage{ragged2e}
\usepackage{enumitem} 
\usepackage{longtable}
\usepackage{adjustbox}

\usepackage{cleveref} 
\usepackage{subfig}
\biboptions{sort&compress} 

\crefname{figure}{}{}
\crefrangelabelformat{figure}{#3#1#4--#5\crefstripprefix{#1}{#2}#6}
\crefmultiformat{figure}{~#2#1\xdef\mycreffirstarg{#1}#3}%
{ and~#2{\crefstripprefix{\mycreffirstarg}{#1}}#3}{, #2#1#3}{ and~#2#1#3}
 
\justifying
\usepackage{physics}
\usepackage[left]{lineno}
\usepackage[none]{hyphenat}
\usepackage[nodisplayskipstretch]{setspace}

\makeatletter
\newcommand\tabfill[1]{%
    \dimen@\linewidth
    \advance\dimen@\@totalleftmargin
    \advance\dimen@-\dimen@curtab
    \parbox[t]\dimen@{#1\ifhmode\strut\fi}
}\makeatother
\begin{document}
\begin{frontmatter}
\title{Empirical modeling and hybrid machine learning framework for nucleate pool boiling on microchannel structured surfaces}

\author{Vijay Kuberan}
\author{Sateesh Gedupudi\corref{CB}}

\ead{sateeshg@iitm.ac.in}

\address{Heat Transfer and Thermal Power Laboratory, Department of Mechanical Engineering, Indian Institute of Technology Madras, Chennai-600036, India}
\cortext[CB]{Associate Professor and Corresponding author. Tel.:+91 44 2257 4721}

\vspace{-2cm}
\begin{abstract}
Micro-structured surfaces influence nucleation characteristics and bubble dynamics besides increasing the heat transfer surface area, thus enabling efficient nucleate boiling heat transfer. Modeling the pool boiling heat transfer characteristics of these surfaces under varied conditions is essential in diverse applications. A new empirical correlation for nucleate boiling on microchannel structured surfaces has been proposed with the data collected from various experiments in previous studies since the existing correlations are limited by their accuracy and narrow operating ranges. This study also examines various Machine Learning (ML) algorithms and Deep Neural Networks (DNN) on the microchannel structured surfaces dataset to predict the nucleate pool boiling Heat Transfer Coefficient (HTC). With the aim to integrate both the ML and domain knowledge, a Physics-Informed Machine Learning Aided Framework (PIMLAF) is proposed. The proposed correlation in this study is employed as the prior physics-based model for PIMLAF, and a DNN is employed to model the residuals of the prior model. This hybrid framework achieved the best performance in comparison to the other ML models and DNNs. This framework is able to generalize well for different datasets because the proposed correlation provides the baseline knowledge of the boiling behavior. Also, SHAP interpretation analysis identifies the critical parameters impacting the model predictions and their effect on HTC prediction. This analysis further makes the model more robust and reliable.
\end{abstract}

\end{frontmatter}


\clearpage
\section*{Nomenclature}
\begin{tabbing}
    xxxxxxxxxxxxxxxxxx \= \kill

    $C_{pl}$ \> Specific heat of the liquid corresponding to $T_{film}$ (kJ/kgK) \\
    $C_{pv}$ \> Specific heat of the vapor corresponding to $T_{sat}$ (kJ/kgK) \\
    $C_{s}$\> Fluid surface coefficient in the Pioro correlation \\
    $C_{sf}$\> Fluid surface coefficient in the Rohsenow correlation \\
    $D_{d}$\> Bubble departure diameter \\
    $D_{h}$\> Hydraulic diameter \\
    $h$ \> Heat transfer coefficient (kW/m$^2$K) \\
    $h_f$ \> Fin width ($\mu$m)  \\
    $h_l$ \> Specific enthalpy of liquid (kJ/kg) \\
    $h_{lv}$ \> Latent heat of vapourisation (kJ/kg) \\
    $h_v$ \> Specific enthalpy of vapor (kJ/kg) \\
    $k_{w}$ \> Thermal conductivity of the substrate (W/mK) \\
    $L_c$ \> Boiling length scale ($\mu$m) \\
    $m$ \> Experimental constant in Rohsenow correlation\\
    $M_{f}$ \> Molecular mass of fluid (kg/kmol) \\
    $M_{w}$ \> Molecular mass of water (kg/kmol) \\
    $n$ \> Constant in Pioro correlation\\
    $p$ \> Pitch ($\mu$m)  \\
    $P_{c}$ \> Critical pressure (bar) \\
    $P_{film}$ \> Saturation pressure corresponding to $T_{film}$ (bar) \\
    $P_{op}$ \> Operating pressure (bar) \\
    $P_{r}$  \> Reduced pressure (bar) \\
    $R^2$ \> Coefficient of determination \\
    $r_{cav}$ \> Cavity radius required for nucleation ($\mu$m) \\
    $R_q$ \> Surface roughness ($\mu$m) \\
    $T_{film}$ \> Film temperature (K) \\
    $T_w$ \> Temperature of the heated surface (K) \\
    $\triangle T$ \> Wall superheat (K) \\
    $w_g$ \> Groove width ($\mu$m)  \\
    $w_f$ \> Fin width ($\mu$m) \\
    \\
    Greek symbols \\
    $\alpha$ \> Thermal diffusivity (m$^2$/s) \\
    $\kappa$ \> Thermal conductivity (W/mK) \\
    $\lambda$ \> Area augmentation factor \\ 
    $\mu$ \> Dynamic viscosity (kg/m s) \\
    $\nu$ \> Kinematic viscosity (m$^2$/s) \\
    $\rho$ \> Density (kg/m$^3$) \\
    $\sigma$ \> Surface tension (N/m) \\
    $\theta$ \> Contact angle (degrees) \\
    \\
    Non-dimensional Quantities \\
    $Pr$ \> Prandtl number \\
    \\
    Abbreviations \\
    HTC \> Heat Transfer Coefficient \\
    MAE \> Mean Absolute Error \\
    ML \> Machine Learning \\
    PIMLAF \> Physics-Informed Machine Learning Aided Framework\\
    RMSE \> Root Mean Squared Error \\
    SD \> Standard Deviation \\
    \\
    Subscripts \\
    corr \> correlation \\
    c \> critical \\
    l \> liquid \\
    sat \> saturated \\
    v \> vapor \\
\end{tabbing}

\clearpage

\section{Introduction}
The two-phase heat transfer process finds wide applications in many industrial domains to improve system efficiency. Boiling is an efficient two-phase heat transfer process that dissipates large thermal energy with a small temperature difference, finding extensive applications in thermal management systems across various domains. Active and passive enhancement methods are employed to boost the boiling heat transfer further \cite{Bergles1997}. Owing to the use of external sources by active methods, passive techniques are more widely adopted than the former to enhance the boiling performance \cite{Li2020}. Surface modification of the substrate by roughening, coating, structuring, patterning, and hybrid surface treatment are the main passive enhancement methods employed extensively, mainly due to their increased nucleation sites, capillary effects, and optimum wettability conditions \cite{MAHMOUD2021101024, MAHMOUD2021101023}. 

\subsection{Boiling on structured surfaces}

\begin{figure}[!h]    
\centering
    \includegraphics[width=18cm]{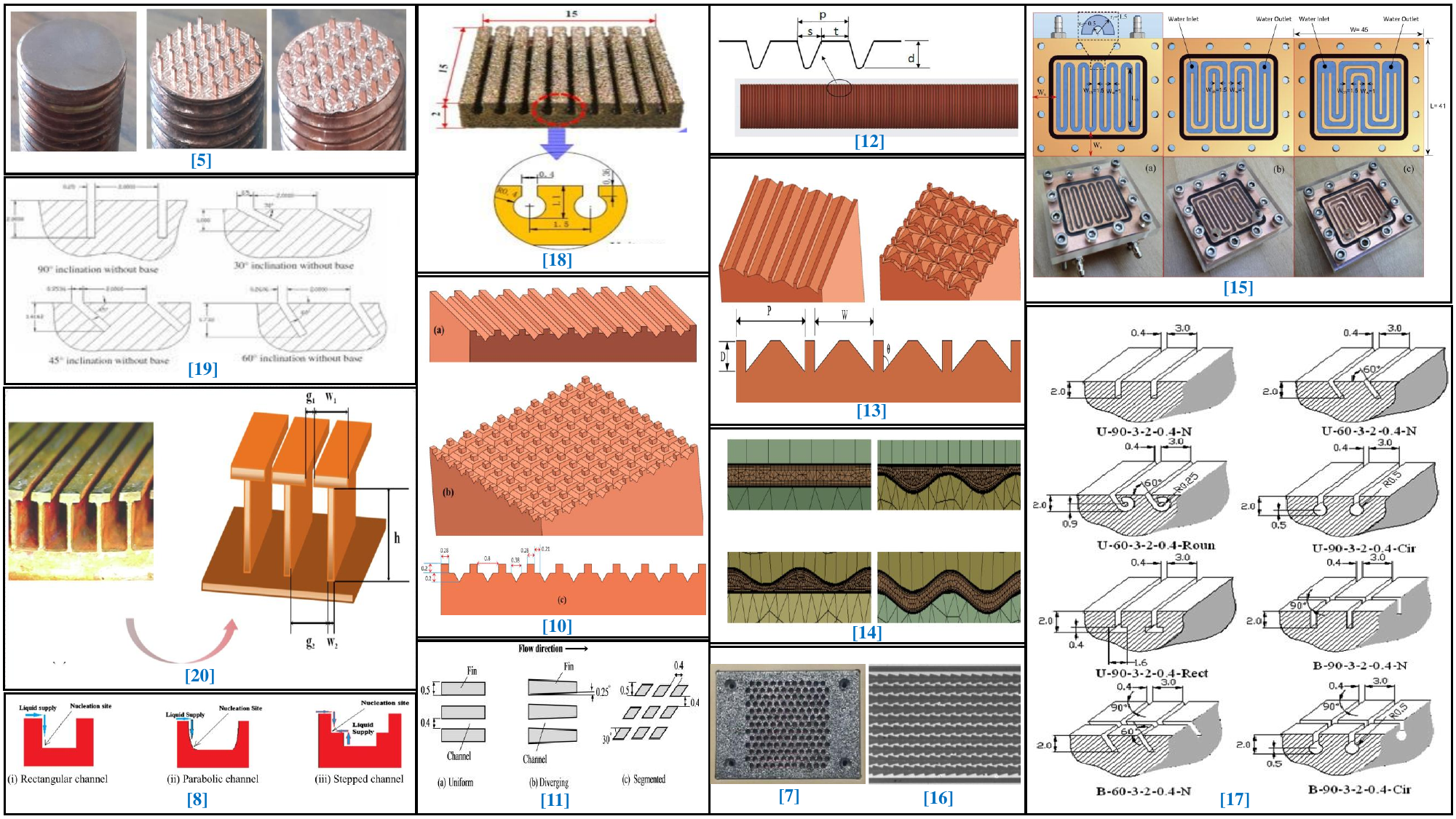}
\caption{Different structured surfaces (reproduced with permissions).}
\label{fig: Different structured surfaces} 
\end{figure}

Structured surfaces with fins and channels of various scales, sizes, and shapes like rectangular fins \cite{JASWAL2023121167}, trapezoidal fins \cite{JASWAL2023121167}, square, triangular, and circular micro-pin-fins \cite{HASAN2014598}, honeycomb structured fins \cite{WANG2020115036}, rectangular channels \cite{WALUNJ2018672, en14113062}, stepped microchannels \cite{WALUNJ2018672, RANJAN2023104351}, parabolic microchannels \cite{WALUNJ2018672}, divergent channels \cite{PRAJAPATI2015711}, V-grooved channels \cite{Balkrushna2022, KUMAR2024125096}, wavy channels \cite{CHENG2020104456}, serpentine channels \cite{ALNEAMA2017709}, zig-zag channels \cite{MA2017157}, inclined channels \cite{DAS20093643}, channels with reentrant cavities \cite{DAS20093643, PI2020119920}, channels with different inclined angle \cite{DAS20101422}, T-shaped micro fins \cite{JIANG2022110663}, and mixed wettability patterns \cite{BETZ2013733, LIM2020119360, rahman2017} augment the boiling heat transfer due to increased effective heat transfer area, improved wicking ability, and creates separate paths for bubble growth and departure \cite{LI2020109926}. Fig. \ref{fig: Different structured surfaces} shows the different kinds of these structured surfaces. Effectively modeling all these surfaces by computational methods to predict the heat transfer characteristics is difficult as there are complex interactions between various parameters. Correlation-based and Machine Learning-based analyses are capable of effectively modeling these surfaces when provided with large datasets. The limited dataset constrains the modeling of boiling on all the types of structured surfaces, as the surface features in each type are distinct and cannot be generalizable for analysis. For the present study, microchannel structured surfaces are analyzed due to the availability of a substantial dataset from the literature. 

Different studies on the microchannel structured surfaces have been carried out extensively by various authors. A study by Kaniowski and Pastuszko \cite{en14113062} examined various configurations having microchannel widths of half the pitch and concluded that the surface extension factor has a substantial effect on the Critical Heat Flux (CHF). A similar study by Orman et al. \cite{en13112700} with water and ethanol identified that deep-grooved geometries and surface enhancement factors significantly impact the boiling performance. An investigation by Cooke and Kandlikar \cite{COOKE20121004} on 10 different geometries of microchannel copper surfaces inferred that wider and deeper channels with thinner fins performed better than others. Kalani and Kandlikar \cite{kalanifc8720212,kalaniethanol20212} conducted experiments with FC-87 and ethanol over microchannel surfaces and determined that increased surface area plays a pivotal role in enhanced performance. Balkrushna et al. \cite{Shah2023} carried out experiments with R123 and R141b at different pressures and identified that microchannel surfaces at higher pressures enhanced boiling Heat Transfer Coefficient (HTC) over plain surfaces. A study by Kwak et al. \cite{KWAK2018111} concluded that high aspect ratio microchannels significantly enhanced the HTC and Critical Heat Flux (CHF) due to their increased capillary wicking capability. To predict the boiling HTC, correlations by Rohsenow \cite{Rohsenow1952}, Pioro \cite{Pioro1999}, Stephan \& Abdelsalam \cite{STEPHAN198073},  Borishansky \cite{borishanskii1969correlation}, Jung \cite{JUNG2003240}, Labuntsov \cite{labuntsov1973heat}, Gorenflo \& Kenning \cite{gorenflo1993vdi}, Kruzhilin \cite{kruzhilin1947free}, Tarrad \& Khudor \cite{HussainTarrad2014ACF}, Cornwell-Housten \cite{Cornwell1994}, Shah \cite{Balkrushna2022},  Cooper \cite{COOPER1984157}, Stephan \& Preusser \cite{stephan1979warmeubergang}, Ribatski \& Jabardo \cite{Ribatski2003}, Kichigin \& Tobilevich \cite{kichigin1955generalization}, and Kutateladze \cite{kutateladze1966concise, kutateladze1990heat} have been proposed. Accurate prediction of HTC on these surfaces is limited by these correlations, as they fail to account for the various influencing parameters and complex interplay of features. The above studies signify the enhanced boiling characteristics on structured surfaces.

\subsection{ML modeling}
Machine Learning techniques can effectively address these challenges by capturing the interactions among features, thus identifying and modeling the non-linear relationship in the data \cite{Sarker2021}. Various domains have employed ML in their analysis due to their significant advantages. A study by Shanmugam and Maganti \cite{SHANMUGAM2024123769} analyzed four different ML algorithms, namely K-Nearest Neighbor (KNN), Light Gradient Boosting Machine (LightGBM), Extreme Gradient Boosting (XGBoost), and Artificial Neural Network (ANN) to predict the thermal performance of microchannel heat sinks under uniform and non-uniform heat loads with 560 data points. Among these, XGBoost and LightGBM perform the best. The author also aims to leverage a diverse and large dataset for model robustness for future research. The above study has been extended to predict the Nusselt number on oblique pin-fin heat sinks \cite{shanmugam2024predicting} using ML models with 893 data points, including KNN, XGBoost, LightGBM, and Random Forest (RF). The findings showed that these models exhibited lower MAE over traditional correlations. Oh and Guo \cite{oh2023prediction} predicted the forced convection nusselt number on microscale pin-fin heat sinks, with ANN achieving the lowest MAPE for different operating conditions and geometries. A similar study by Traverso et al. \cite{traverso2023machine} adopted Gaussian process regression (GPR) to predict the HTC in microchannels using Brunel two-phase flow dataset. Loyola-Fuentes et al. \cite{loyola2023application} explored RF and Deep Neural Networks (DNN) to estimate the condensation HTC in micro fin tubes (4333 data points). The author analyzed the impact of input variables and compared the model performance with non-dimensional input parameters. Sei et al. \cite{seiprediction} integrated DNN and GPR to predict HTC and uncertainties in predicted HTC in horizontal flow boiling within mini channels. Huang et al. \cite{en17010044} employed ANN to optimize the microchannel geometries based on pressure drop, HTC, and refrigerant convection thermal resistance. A study by Qiu et al. \cite{QIU2021121607} predicted the pressure drop in mini and microchannels using ANN, LightGBM, XGBoost, and KNN using different combinations of 23 dimensionless input parameters using 2787 data points under different conditions. It showed better prediction than widely used correlations in flow boiling. To predict the pressure drop in the two-phase flow of R407C in a horizontal copper tube, Khosravi et al. \cite{Khosravi2018} used 500 experimental data and inferred that ANN and Group Method of Data Handling (GMDH) type neural network performed better with 99\% accuracy than Support Vector Regression (SVR) model. 

A  study by Alizadehdakhel et al. \cite{ALIZADEHDAKHEL2009850} compared the results of ANN and CFD predictions of pressure drop and two-phase flow regimes in a tube and concluded that the ANN model accuracy can be improved with 443 datasets. A similar study by Bar et al. \cite{BAR2013813} on air-water flow through U-bends observed that the ANN model accurately predicts the two-phase frictional pressure drop when modeled using 241 experimental data. Azizi and Ahmadloo \cite{AZIZI2016203} predicted the condensation convective HTC in inclined tubes with R134a as the working fluid under varying operating conditions (440 data points) with a $R^2$ value of 0.995 by implementing the ANN model. A study by Vijay and Gedupudi \cite{K2024, kuberan2024} predicted the nucleate pool boiling HTC on plain, roughened, thin-film coated, and porous-coated substrates. The XGBoost and Categorical Boosting (CatBoost) models outperformed other ML models in the HTC prediction. In a study by Aliyu et al.\cite{ALIYU2023104452}, the entrained liquid fraction in annular gas-liquid two-phase flows has been predicted accurately with a $R^2$ value of 0.98 by the ANN model with 1367 data points, with superficial gas velocity having a major impact on model prediction. Arteaga et al. \cite{arteaga2021machine} classified the two-phase gas-liquid flow regimes employing the Extra Trees (ET) regressor and ANN model utilizing 9029 data points. Here, the ET regressor model achieves an accuracy of 98\%, surpassing ANN performance. Jalili and Mahmoudi \cite{JALILI2025126680} applied the Physics-Informed Neural Networks (PINNs) to model the film boiling heat transfer process with 776 data generated from CFD simulations. A similar study by Kim et al. \cite{KIM2022122839}  employed a hybrid technique of incorporating the Groeneveld- Stewart correlation with ML models (RF, ANN) using 400 data points to predict minimum film boiling temperature. This Physics-Informed Machine Learning-Aided Framework (PIMLAF) exhibited better performance than individual ML models. The above studies highlight the adoption of ML in diverse two-phase flow applications. 

\subsection{Objective and significance of the present study} \label{sec: Objective of the present study}
Various parameters inherently govern the boiling heat transfer process, and a complete understanding of this phenomenon is limited by the complex dynamics of various features. Modeling and predicting heat transfer characteristics by computational methods and empirical correlations are limited to specific operation conditions and reduced accuracy. There is a scarcity of considerable data to predict the boiling behavior on other structured surfaces. While sufficient data exists for microchannel structured surfaces, a reliable and accurate model to predict the nucleate pool boiling heat transfer performance on microchannel structured surfaces has still not been developed.

The current study aims to develop a model for the prediction of the boiling HTC on microchannel structured surfaces through empirical, ML, and hybrid framework approaches. This study develops an ML model to predict boiling HTC with a large dataset (7128 data points) amassed from different studies under diverse conditions. Also, comprehensive and prominent parameters affecting the boiling behavior in microchannels are selected as the input for the model. Moreover, liquid thermophysical properties are calculated at the surface film temperature to accurately model the heat transfer at the liquid-vapor interface. Furthermore, the study seeks to employ a hybrid ML framework, integrating the developed empirical correlation and ML model to improve the prediction accuracy. The study also aims to understand the impact of input parameters on model prediction using SHAP (Shapley Additive exPlanations) analysis. The study presents the hybrid ML analysis for water and other fluids separately. Thus, this research study aims to develop a robust, reliable, and accurate model for nucleate pool boiling on microchannel structured surfaces, which is capable of performing better across a wide range of parameters.

\section{Methodology}\label{sec: Methodology}

\begin{figure}[!h]    
\centering
    \includegraphics[width=18cm]{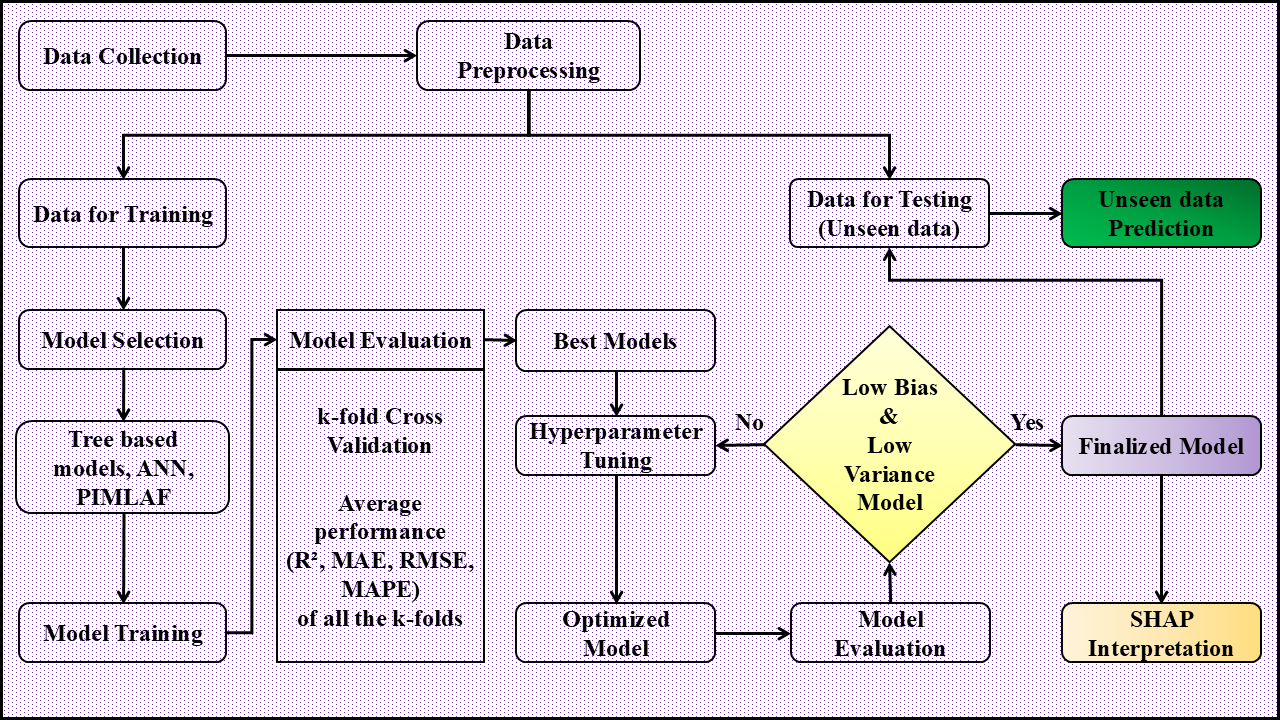}
\caption{Overall methodology.}
\label{fig: Methodology}
\end{figure}

The ML analysis involves data collection, feature selection, data preprocessing, selection of ML models, model training, model optimization through hyperparameter tuning, model evaluation through k-fold cross-validation, model testing, and model interpretation through SHAP analysis. Fig. \ref{fig: Methodology} shows the overall methodology implemented in this analysis. 

\subsection{Data collection}
A wide range of datasets on pool boiling of microchannel structured surfaces has been collected from various research studies \cite{WALUNJ2018672, CHEN2022106298, en14113062, rahman2017, COOKE20121004, kalanifc8720212, kalaniethanol20212, Shah2023, en13112700, KWAK2018111}. The data include a diverse set of conditions, including water, R-141b, R-123, and ethanol as the working fluids; copper and silicon wafers as the substrate material; different operating pressures; and multiple microchannel structured surface configurations, all at saturation conditions. A total of 7128 data points have been compiled to predict the heat transfer coefficient.

\subsection{Feature identification}

\begin{figure}[H]    
\centering
    \includegraphics[width=18cm, height=9cm]{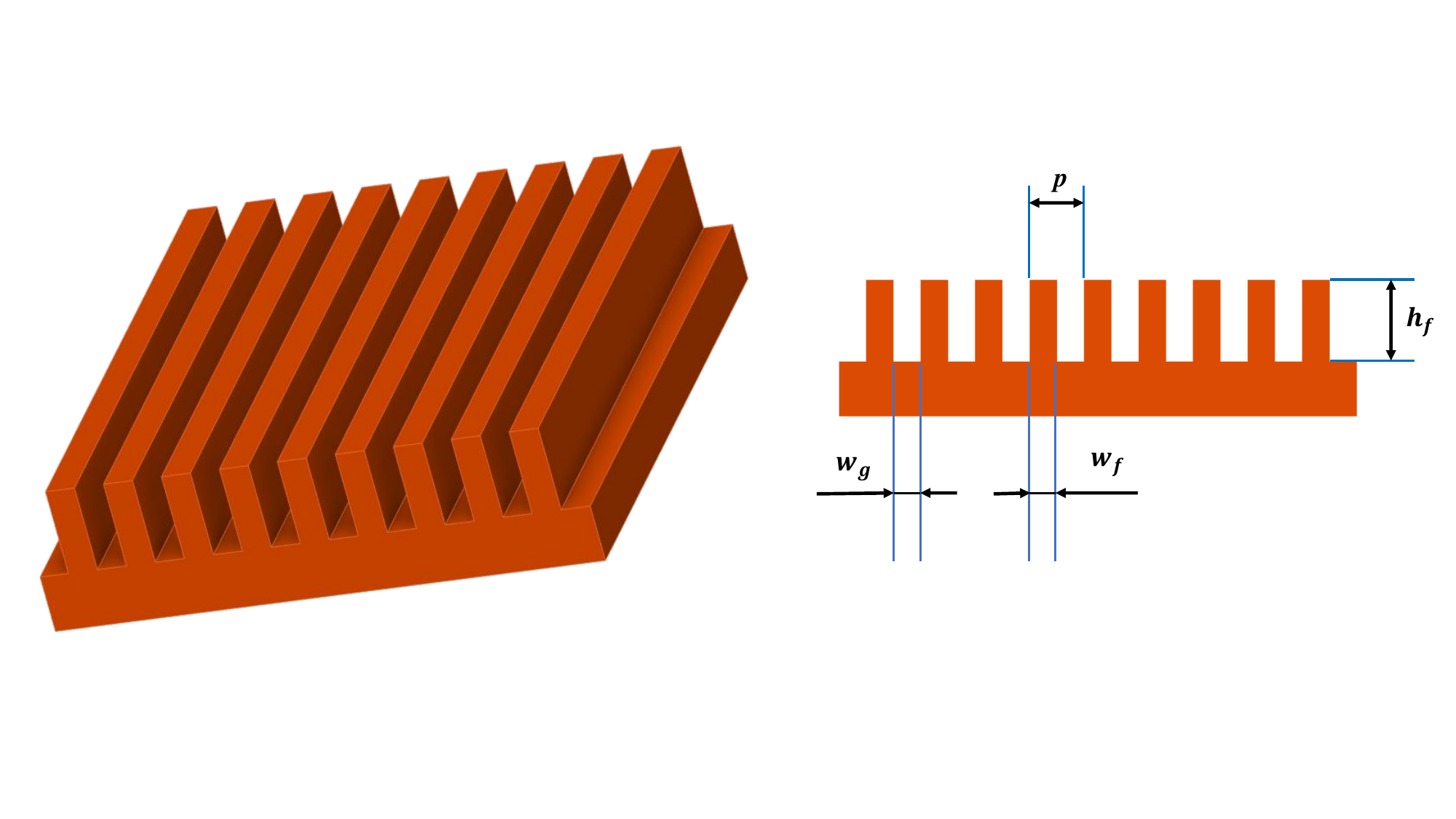}
\caption{Surface characteristics of a microchannel structured surface.}
\label{fig: microchannels_nomenclature}
\end{figure}

Appropriate selection of input features/parameters helps the model to understand the complex relationship among variables and capture the boiling behavior in microchannels. Hence, this study considers the significant parameters, including surface properties ($k_{w}$, $R_q$, $\theta$, $w_g$, $w_f$, $h_f$, $p$, $\lambda$), operating conditions($\triangle T$, $T_w$, $P_{op}$, $M_f$, $T_{sat}$), thermophysical properties ($\rho_l$, $\rho_v$,  $C_{pl}$, $C_{pv}$, $\mu_l$, $\mu_v$,  $k_l$, $k_v$, $\sigma$, $h_{lv}$), and $P_{film}$ (saturation pressure corresponding to film temperature). Also, liquid thermophysical properties are determined at film temperature ( $T_{film}$ = ($T_{w}$ + $T_{sat}$) / 2) to capture the thermal interactions at the vapor-liquid interface \cite{kuberan2024}. As the difference in pressure inside the vapor bubbles and the system pressure is negligible, vapor thermophysical properties are taken at saturation temperature ($T_{sat}$). Here, $P_{op}$ is the operating pressure/system pressure, and $P_{film}$ is the saturation temperature corresponding to the film temperature at the liquid-vapor interface.
The microchannel structured surface parameters are shown in Fig. \ref{fig: microchannels_nomenclature}.

Tables \ref{tab: Range of parameters for microchannels} and \ref{tab: Features and their expressions} represent the range of parameters for the microchannel structured surfaces dataset and various expressions used in this study. The fluid properties are sourced from the CoolProp \cite{Bell2014} and National Institute of Standards and Technology (NIST) \cite{lemmon_fluid_systems_chemistry_webbook} databases.

\begin{table}[!h]
\centering
\caption{Range of parameters and target variable ($h$) for microchannel structured surfaces.}
\label{tab: Range of parameters for microchannels}
\begin{adjustbox}{max width=\textwidth}
\begin{tabular}{@{}lllll@{}}
\toprule
\textbf{Features} & \textbf{Min} & \textbf{Max} & \textbf{Mean} & \textbf{SD} \\ 
\midrule
$\triangle T$ (K) & 1.19 & 52.65 & 10.19 & 8.35 \\
$T_w$ (K) & 304.06 & 425.80 & 342.23 & 27.82 \\
$P_{op}$ (bar) & 0.17 & 1.75 & 1.03 & 0.44 \\
$T_{sat}$ (K) & 300.95 & 373.15 & 332.04 & 24.78 \\
$k_w$ (W/mK) & 130 & 401 & 387.43 & 59.11 \\
$R_q$ ($\mu$m) & 0.12 & 6.40 & 1.53 & 1.91 \\
$M_f$ (kg/kmol) & 18.02 & 152.93 & 79.58 & 50.71 \\
$w_g$ ($\mu$m) & 30 & 1150 & 432.58 & 254.74 \\
$w_f$ ($\mu$m) & 30 & 1100 & 406.09 & 245.16 \\
$h_f$ ($\mu$m) & 10 & 600 & 259.64 & 153.69 \\
$p$ ($\mu$m) & 60 & 2250 & 838.67 & 460.19 \\
$\lambda$ & 1.19 & 3.50 & 1.78 & 0.60 \\
$\theta$$^\circ$ & 5 & 106 & 18.09 & 28.18 \\
$P_{film}$ (bar) & 0.17 & 2.42 & 1.20 & 0.45 \\
$\rho_l$ (kg/m$^3$) & 724.60 & 1452.69 & 1029.95 & 268.72 \\
$C_{pl}$ (kJ/kgK) & 1.02 & 4.25 & 2.34 & 1.19 \\
$\mu_l$ (kg m$^{-1}$s$^{-1}$) & $2.20 \times 10^{-4}$ & $8.30 \times 10^{-4}$ & $3.95 \times 10^{-4}$ & $1.24 \times 10^{-4}$ \\
$k_l$ (W/mK) & 0.070 & 0.683 & 0.218 & 0.221 \\
$\sigma$ (N/m) & $1.26 \times 10^{-2}$ & $5.87 \times 10^{-2}$ & $2.37 \times 10^{-2}$ & $1.62 \times 10^{-2}$ \\
$\rho_v$ (kg/m$^3$) & 0.29 & 10.84 & 3.78 & 3.50 \\
$C_{pv}$ (kJ/kgK) & 0.70 & 2.08 & 1.34 & 0.52 \\
$\mu_v$ (kg m$^{-1}$s$^{-1}$) & $9.0 \times 10^{-6}$ & $1.20 \times 10^{-5}$ & $1.10 \times 10^{-5}$ & $1.0 \times 10^{-6}$ \\
$k_v$ (W/mK) & 0.009 & 0.025 & 0.016 & 0.006 \\
$h_{lv}$ (kJ/kg) & 163.24 & 2256.40 & 828.05 & 742.22 \\
$h$ (kW/m$^2$K) & 0.87 & 390.24 & 23.22 & 35.59 \\

\bottomrule
\end{tabular}
\end{adjustbox}
\end{table}

\begin{table}[!h]
\centering
\caption{Important expressions used in the analysis.}
\label{tab: Features and their expressions}
\begin{adjustbox}{max width=\textwidth}
\begin{tabular}{ll}
\toprule
\hspace{10pt}\textbf{Features} & \hspace{10pt}\textbf{Expressions} \\
\midrule
\hspace{10pt}$\Delta T$ & \hspace{10pt} $T_w - T_{sat}$ \\ [4pt]
\hspace{10pt}$Pr_l$ &  \hspace{10pt} $\frac{\mu_{l} \cdot C_{pl}}{k_l}$ \\ [4pt]
\hspace{10pt}$r_{cav}$ & \hspace{10pt}  $\frac{2\sigma\left(\frac{1}{\rho_v} - \frac{1}{\rho_l}\right)T_{sat}}{\Delta T \cdot h_{lv}}$ \cite{MAHMOUD2021101024} \\ [4pt]
\hspace{10pt}$P_{r}$ & \hspace{10pt}  $\frac{P_{op}}{P_{c}}$ \\ [4pt]
\hspace{10pt}$T_{r}$ & \hspace{10pt}  $\frac{T_{sat}}{T_{c}}$ \\ [4pt]
\hspace{10pt}$L_c$ &  \hspace{10pt} $\sqrt{\frac{\sigma}{g(\rho_l - \rho_v)}}$ \cite{Pioro1999} \\ [9pt]
\hspace{10pt}$D_d$ &  \hspace{10pt} 0.0208 $\cdot \theta \cdot \sqrt{\frac{\sigma}{g (\rho_l - \rho_g)}}$ \cite{fritz1935berechnung}\\ [9pt]
\bottomrule
\end{tabular}
\end{adjustbox}
\end{table}

\subsection{Data visualization}

\begin{figure}[H]    
\centering
    \includegraphics[width=18cm, height = 24cm]{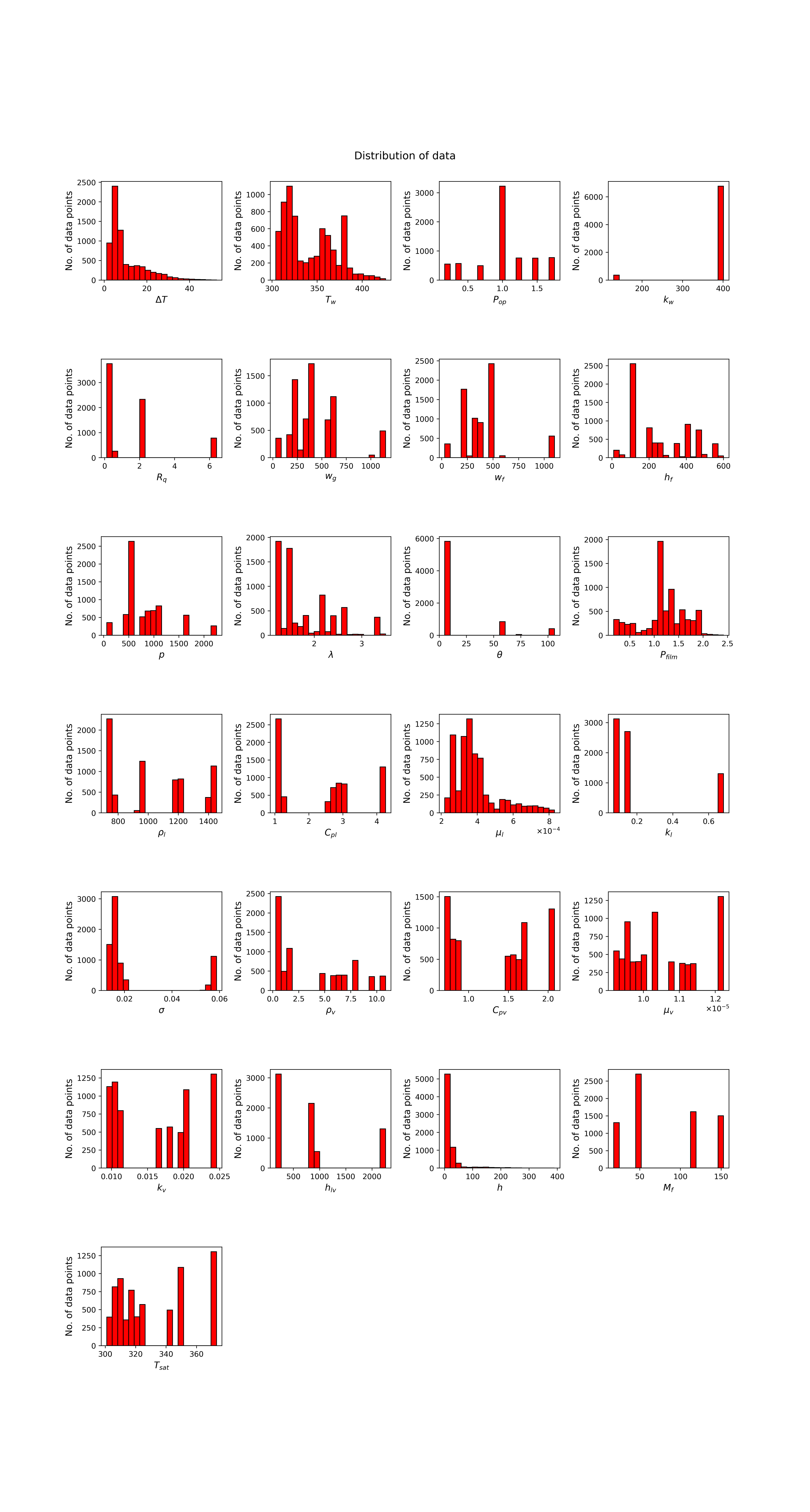}
\caption{Data distribution of microchannel structured surfaces data.}
\label{fig: histogram_microchannels}
\end{figure}

\begin{figure}[H]    
\centering
    \includegraphics[width=18cm, height = 12cm] {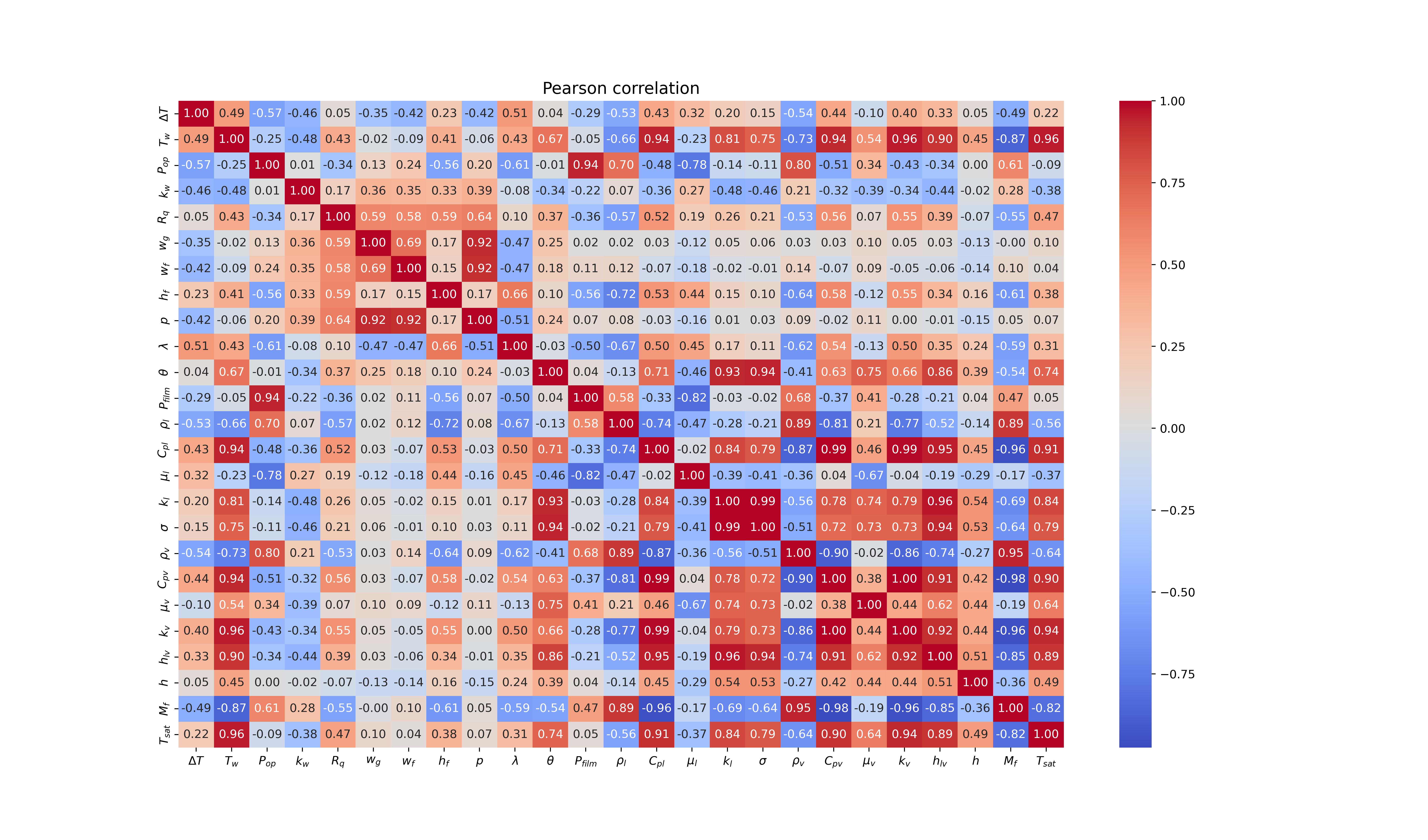}
\caption{Pearson correlation chart for microchannel structured surfaces data.}
\label{fig: pearson_microchannels}
\end{figure}

\begin{figure}[H]    
\centering
    \includegraphics[width=18cm, height=12cm]{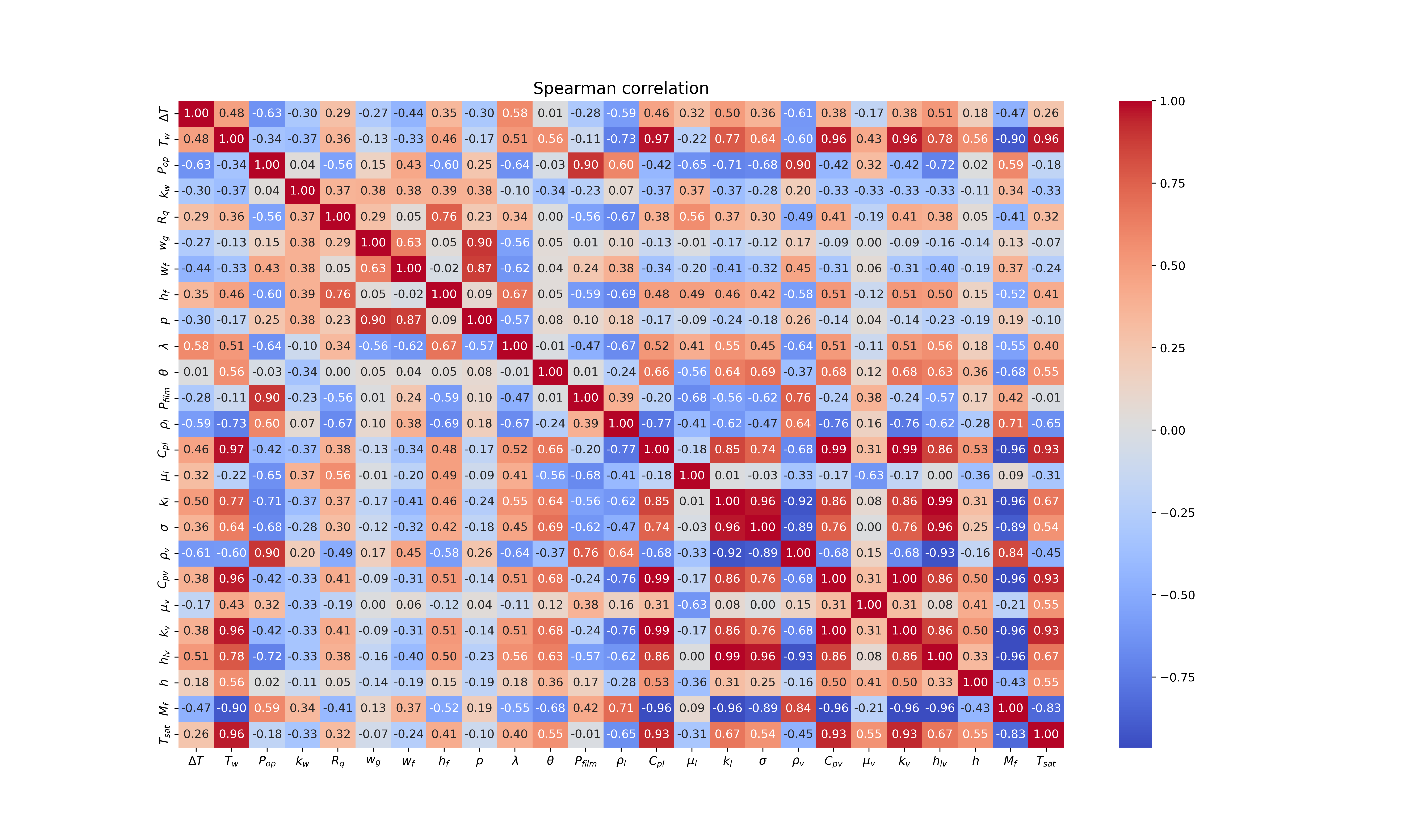}
\caption{Spearman correlation chart for microchannel structured surfaces data.}
\label{fig: spearman_microchannels}
\end{figure}

The data distribution, Pearson correlation chart, and Spearman correlation chart of the microchannel structured surfaces data are presented in Figs. \ref{fig: histogram_microchannels}, \ref{fig: pearson_microchannels}, and \ref{fig: spearman_microchannels} respectively. Values of correlation coefficients close to +1 are indicated in dark blue and in dark red for values close to -1. Pearson explains only the linear correlation between the variables, whereas Spearman correlation represents the monotonic relationship. They do not capture non-linear relationships. Hence, variables having no linear/monotonic relationship and having non-linear relationships have correlation coefficients of nearly zero.

\subsection{Data preprocessing}
It is important to feed quality data into the ML model. The collected data must be processed carefully to identify the missing data, duplicates, and outliers. Handling missing data in Machine Learning analysis is very crucial. In this analysis, missing data have been imputed using LightGBM \cite{NIPS2017_6449f44a}. The percentage of data imputed in this analysis is 20\%. Also, the duplicates and outliers were removed, and 7128 data points were finally used for analysis. 

Here, one-hot encoding is performed to convert the categorical variable into numerical format. The categorical variable - Fluid is one-hot encoded. Z-score standardization establishes equal feature contribution during model training by transforming every feature to have a mean of 0 and unit variance when handling data with different scales \cite{Ahsan2021}. 

\clearpage
\subsection{Machine learning models}
Various ML models are employed in this analysis. The top-performing models are described in this section. 

\subsubsection{Deep Neural Networks}
\begin{figure}[H]    
\centering
    \includegraphics[width=18cm]{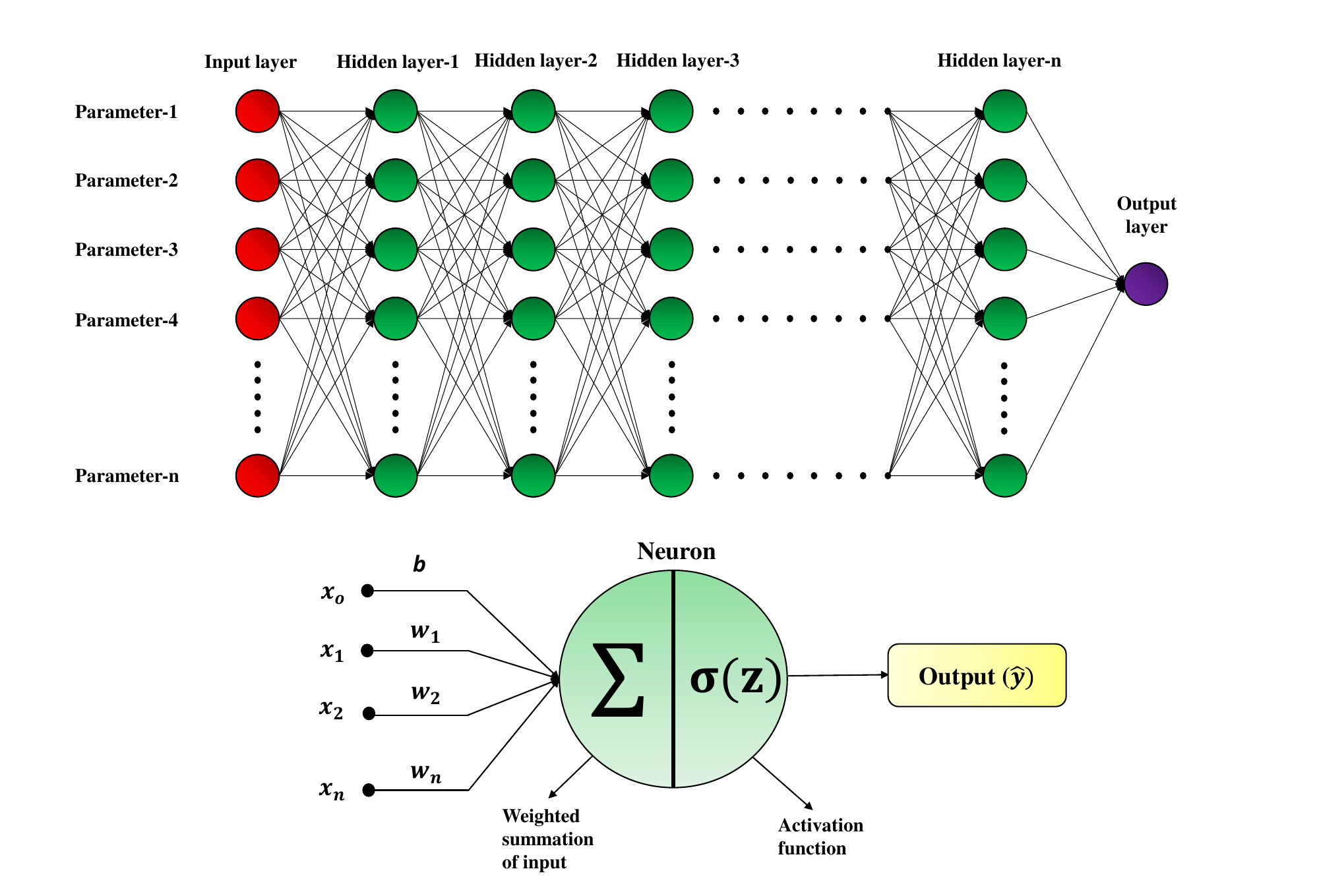}
\caption{Deep Neural Networks.}
\label{fig: DNN_model}
\end{figure}

Deep neural network (DNN) is a powerful ML technique and has extensive applications in various sectors ranging from regression, Computer Vision (CV), and Natural Language Processing (NLP) to Generative AI (GenAI) \cite{SCHMIDHUBER201585, bishop2006pattern}. Their success lies in modeling complex non-linear relationships in high-dimensional data. This ability is due to the presence of neurons, which is the building block of the neural network. Figure \ref{fig: DNN_model} shows the internal structure of a neuron. They read the input, perform a weighted combination (z) on the inputs using weights (w) and biases(b), and subsequently pass through the activation function. The activation function is responsible for introducing the non-linearity in the data. Table \ref{tab: Activation Functions} represents various activation functions and their formulations. In this analysis, the ELU (Exponential Linear Unit) activation function is used. Then, the output of the activation function is passed to the successive neurons in the next layer. This process is repeated across various hidden layers. Generally, a neural network model consists of an input layer, hidden layers, and an output layer, as depicted in Figure \ref{fig: DNN_model}. The input layer reads the input, where the number of features and number of training data represents the size of this layer. Then, it is processed through the hidden layers, where the patterns and complex relationships in the data are captured through weights, biases, and activation functions. Finally, the output layer outputs the model prediction. The use of many hidden layers and neurons improves the model's accuracy and generalizability but should be optimized to reduce overfitting. The optimum architecture is developed based on the nature of the data. This is the forward propagation.

\begin{table}[H]
\centering
\caption{Commonly used activation functions.}
\label{tab: Activation Functions}
\begin{adjustbox}{max width=\textwidth}
\begin{tabular}{@{}ll@{}}
\toprule
\textbf{Activation Function} & \textbf{Formula}                                                   \\ 
\midrule
Sigmoid \cite{cybenko1988continuous}            & \( \displaystyle \sigma(x) = \frac{1}{1 + e^{-x}} \)                       \\
Tanh \cite{chentanh}             & \( \displaystyle \text{tanh}(x) = \frac{e^x - e^{-x}}{e^x + e^{-x}} \)     \\
ReLU (Rectified Linear Unit) \cite{agarap2019deeplearningusingrectified} & \( \displaystyle \text{ReLU}(x) = \max(0, x) \)                              \\
Leaky ReLU \cite{info12120513}        & \( \displaystyle \text{Leaky ReLU}(x) = \begin{cases} x & \text{if } x > 0 \\ \alpha x & \text{if } x \leq 0 \end{cases} \) \\
ELU (Exponential Linear Unit) \cite{clevert2016fastaccuratedeepnetwork} & \( \displaystyle \text{ELU}(x) = \begin{cases} x & \text{if } x > 0 \\ \alpha (e^x - 1) & \text{if } x \leq 0 \end{cases} \) \\
\bottomrule
\end{tabular}
\end{adjustbox}
\end{table}

The neural network parameters - weights and biases have to be learned and updated to increase the model accuracy through backpropagation. During a single forward pass, the model predictions are compared against the actual values using a loss function. The gradients of the loss function are calculated with respect to the weights and biases and are updated using an optimizer (Gradient descent \cite{ruder2017overviewgradientdescentoptimization}, Adagrad \cite{JMLR:v12:duchi11a}, Adadelta \cite{zeiler2012adadeltaadaptivelearningrate}, Adam \cite{kingma2017adammethodstochasticoptimization}, etc.). The updated weights and biases are then used to do the forward pass and compute the model prediction. The loss function for the new predictions is calculated, and this iterative process is carried out till convergence, where the loss function is the least. This is the backpropagation algorithm. In the loss function, in addition to MSE (Mean Squared Error) loss, Lasso (L1) \cite{10.1111/j.2517-6161.1996.tb02080.x} and Ridge (L2) \cite{Hoerl01021970} regularization terms are added to prevent overfitting. This is called the regularized loss function. L1 regularization simplifies the model by penalizing the weights to zero, while L2 regularization stabilizes the model by penalizing the weights but does not shrink them to zero. The model architecture is finalized after optimizing the hyperparameters and evaluated using various performance metrics. Finally, it can be deployed for real-world predictions.

The entire formulation of a neural network model is described below:

\textbf{1. Neural network parameters initialization:}

For each layer \( l = 1, 2, \dots, L \): 
\[
\mathbf{W}^{(l)}, \mathbf{b}^{(l)} \leftarrow \text{random initialization}
\]

\(\mathbf{W}^{(l)}\) - Weight matrix for layer \(l\), 
\(\mathbf{b}^{(l)}\) - Bias vector for layer \(l\)

\textbf{2. Forward propagation:}

For each layer \( l = 1, 2, \dots, L \):
\[
\mathbf{z}^{(l)} = \mathbf{W}^{(l)} \mathbf{a}^{(l-1)} + \mathbf{b}^{(l)}
\]
\[
\mathbf{a}^{(l)} = \sigma(\mathbf{z}^{(l)})
\]
\[
\hat{\mathbf{y}} = \mathbf{a}^{(L)} = \mathbf{W}^{(L)} \mathbf{a}^{(L-1)} + \mathbf{b}^{(L)}
\]

\(\mathbf{z}^{(l)}\) - Linear combination of inputs, \(\sigma(z)\) - Activation function, 

\(\mathbf{a}^{(l)}\) - Activation for layer \(l\), \(\hat{y}\) - Predicted output

\textbf{3. Loss computation:}

For each data point \( i \):
\[
\mathcal{L} = \underbrace{\frac{1}{N} \sum_{i=1}^{N} (y_i - \hat{y}_i)^2}_{\text{MSE Loss}} + \underbrace{\lambda_1 \sum_{l=1}^{L} \| \mathbf{W}^{(l)} \|_1}_{\text{L1 Loss}} + \underbrace{\lambda_2 \sum_{l=1}^{L} \| \mathbf{W}^{(l)} \|_2^2}_{\text{L2 Loss}}
\]

\(\mathcal{L}\) - Loss function
\(\lambda_1\) - L1 loss regularization parameter, 
\(\lambda_2\) - L2 loss regularization parameter

\textbf{4. Backward propagation:}

For the output layer \( L \):
\[
\frac{\partial \mathcal{L}}{\partial \mathbf{z}^{(L)}} = \frac{\partial \mathcal{L}}{\partial \hat{\mathbf{y}}} \odot \sigma'(\mathbf{z}^{(L)})
\]

For each layer \( l = L, L-1, \dots, 1 \):
\[
\frac{\partial \mathcal{L}}{\partial \mathbf{W}^{(l)}} = \frac{1}{N} \left( \frac{\partial \mathcal{L}}{\partial \mathbf{z}^{(l)}} \cdot (\mathbf{a}^{(l-1)})^T \right) + 2 \lambda_2 \mathbf{W}^{(l)} + \lambda_1 \text{sign}(\mathbf{W}^{(l)})
\]
\[
\frac{\partial \mathcal{L}}{\partial \mathbf{b}^{(l)}} = \frac{1}{N} \sum \frac{\partial \mathcal{L}}{\partial \mathbf{z}^{(l)}}
\]
\[
\frac{\partial \mathcal{L}}{\partial \mathbf{z}^{(l-1)}} = (\mathbf{W}^{(l)})^T \frac{\partial \mathcal{L}}{\partial \mathbf{z}^{(l)}} \odot \sigma'(\mathbf{z}^{(l-1)})
\]

\textbf{5. Parameter update:}

For each layer \( l = 1, 2, \dots, L \):
\[
\mathbf{W}^{(l)} \leftarrow \mathbf{W}^{(l)} - \eta \frac{\partial \mathcal{L}}{\partial \mathbf{W}^{(l)}}
\]
\[
\mathbf{b}^{(l)} \leftarrow \mathbf{b}^{(l)} - \eta \frac{\partial \mathcal{L}}{\partial \mathbf{b}^{(l)}}
\]

\(\eta\) - Learning rate

\textbf{6. Model evaluation on the test data:}

For each test data point \( \mathbf{x}_i \):
\[
\mathbf{z}^{(l)} = \mathbf{W}^{(l)} \mathbf{a}^{(l-1)} + \mathbf{b}^{(l)} \quad \text{for} \quad l = 1, 2, \dots, L
\]
\[
\mathbf{a}^{(l)} = \sigma(\mathbf{z}^{(l)}) \quad \text{for} \quad l = 1, 2, \dots, L-1
\]
\[
\hat{\mathbf{y}}_i = \mathbf{a}^{(L)} \quad \text{(final predicted output)}
\] 

\textbf{7. Model Deployment:} 
The model is deployed to make predictions on new data.

\subsubsection{Physics-Informed Machine Learning Aided Framework (PIMLAF)}
\begin{figure}[H]    
\centering
    \includegraphics[height=9.5cm, width=18cm]{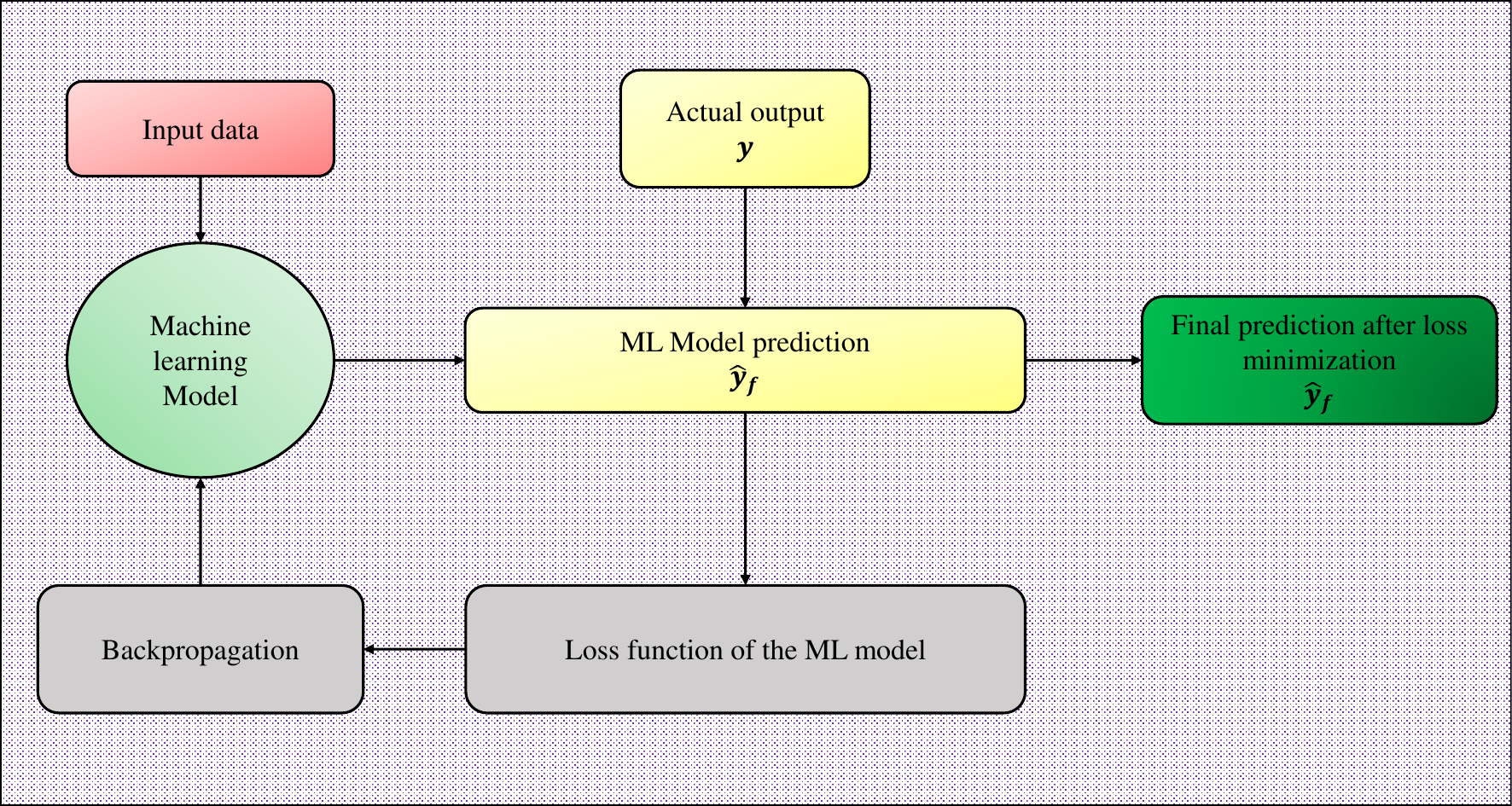}
\caption{Conventional ML model framework.}
\label{fig: conventional_ML_framework}
\end{figure}

\begin{figure}[H]    
\centering
    \includegraphics[width=18cm]{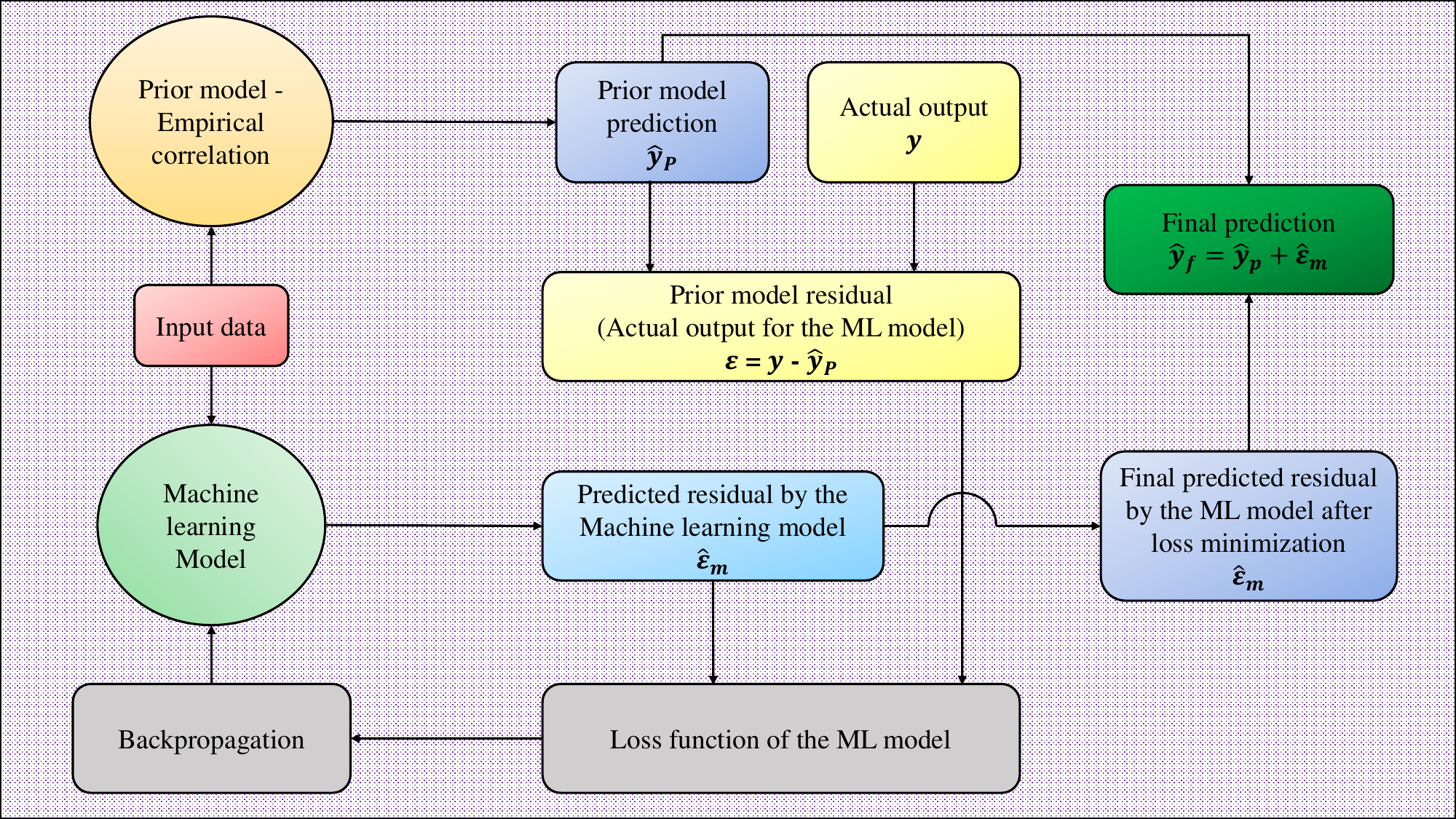}
\caption{PIMLAF framework.}
\label{fig: PIMLAF_framework}
\end{figure}

The reliability of the machine learning model is a critical aspect to be considered while developing the model. The PIMLAF \cite{KIM2022122839, psichogios1992hybrid, SU1992327, thompson1994modeling, ACUNA1999S561} combines the advantages of domain knowledge and ML techniques. Instead of modeling from scratch, this hybrid framework focuses on learning the residuals of the initial prediction from the prior baseline model. The prior model employed is the empirical correlation proposed in this study. This physics-based correlation is able to provide a strong foundation for model prediction.

The hybrid framework initially predicts the output ($\hat{y}_{p}$) from the prior model (physics-based correlation). Then, the predictions are compared against the ground truth values ($y$), and the residual ($\epsilon$ = $y$ - $\hat{y}_{p}$) is calculated. Then, an ML algorithm is trained to predict these residuals. The predicted residual from the ML model is $\hat\epsilon_{m}$. The final prediction ($\hat{y}_{f}$) is the summation of the initial prediction ($\hat{y}_{p}$) from the empirical correlation and the predicted residual ($\hat\epsilon_{m}$) from the ML model (i.e) $\hat{y}_{f}$ = $\hat{y}_{p}$ + $\hat\epsilon_{m}$. Fig. \ref{fig: PIMLAF_framework} depicts the working principle of PIMLAF. Meanwhile, in a conventional ML model (Fig. \ref{fig: conventional_ML_framework}), the output variable is directly predicted.

This approach increases the model's accuracy and reduces the risk of overfitting, as the physics-based correlation improves the generalizability of the ML model. This further enhances the model's reliability on unseen datasets.

\subsubsection{K-nearest neighbor regressor}
KNN \cite{fix1985discriminatory, coverknn} is a simpler algorithm and predicts the output by taking the mean of all the k closest points in the parameter space. The nearest points are calculated by the Euclidean distance. It is easier to implement and doesn't assume any distribution of the existing data.

\subsubsection{Random forest algorithm}
The Random forest model \cite{Breiman2001} works by creating multiple decision trees and averaging the predictions by each decision tree. Each tree takes a random subset of data, and each node selects a random set of features during training to introduce randomness. This reduces overfitting and interprets high-dimensional datasets effectively.

\subsubsection{Extra trees regressor}
ET model \cite{Geurts2006} is analogous to the random forest, but ET introduces additional randomness while creating each tree. RF splits each node based on specific criteria (e.g., information gain, gini impurity) to find the best split, but ET splits each node randomly, and hence, the training is faster.

\subsubsection{XGBoost model}
XGBoost \cite{Chen2016} works on the principle of gradient boosting by building trees in sequence and correcting the errors made by the previous trees. The regularization techniques used in the XGBoost model reduce variance in the model predictions, thus improving the model's generalizability.

\subsubsection{LightGBM model}
LightGBM \cite{NIPS2017_6449f44a} is memory-efficient and trains faster due to the adoption of the histogram-based technique. It is also a boosting algorithm and works well with high-dimensional data. However, proper parameter fine-tuning should be carried out when dealing with limited datasets.

\subsection{Hyperparameter tuning}
Hyperparameters are the parameters defined earlier in a model's training, for e.g., the number of hidden layers, number of neurons in each layer, learning rate, number of epochs during training, and the maximum depth of the tree. They impact the model architecture, improve the model's accuracy, prevent overfitting, and are optimized external to the model training. This analysis adopts random search optimization \cite{bergstra2012random}, which randomly samples over the hyperparameter space, finds better solutions, and is highly efficient in a large dimensional domain.

\subsection{k-fold cross validation}
Cross-validation provides a robust estimation of the model's performance and mitigates overfitting. The procedure involves randomly splitting the training data into k-folds (5-folds in this analysis).  For k iterations, k-1 folds are used for training, and the other fold is used for testing. Each of these k-folds will be a testing set in only one of the iterations. The average and individual performance over all the iterations signifies the model's generalizability over different subsets of the data, and the hyperparameters are optimized for the best performance across all the k-folds.

\subsection{Performance indicators for the model evaluation}
The evaluation/performance indicators represent the model's performance. Table \ref{tab: Performance indicators for the model evaluation} shows the performance metrics (MAE, RMSE, and $R^2$) used in this analysis to evaluate the models.

\begin{table}[H]
\centering
\caption{Performance indicators for the model evaluation.}
\label{tab: Performance indicators for the model evaluation}
\begin{adjustbox}{max width=\textwidth}
\begin{tabular}{@{}p{6.5cm}lp{7cm}@{}}
\toprule
\textbf{Performance metric} & \textbf{Formula} & \textbf{Remarks} \\ 
\midrule
Mean Absolute Error (MAE) &  MAE = \( \displaystyle\frac{1}{n} \sum_{i=1}^{n} |y_i - \hat{y}_i| \)  & Indicates the average absolute error between actual and predicted output. \\
\\
Root Mean Squared Error (RMSE) &  RMSE = \( \displaystyle\sqrt{ \frac{1}{n} \sum_{i=1}^{n} (y_i - \hat{y}_i)^2 } \) & Measures the square root of the mean of squared errors - A error metric in the same scale of the output variable. \\
\\
Coefficient of determination (\( R^2 \)) & \( \displaystyle R^2 = 1 - \frac{\sum\limits_{i=1}^{n} (y_i - \hat{y}_i)^2}{\sum\limits_{i=1}^{n} (y_i - \bar{y})^2} \) & Represents how effectively the model explains the degree of variance in the target variable. \\ &&\( R^2 \) $\to$ 1, \text{ Better performance.} \\&&\( R^2 \) $\to$ 0, \( R^2 \) $<$ 0, \text{Poor performance.} \\
\bottomrule
\end{tabular}
\end{adjustbox}
\end{table}

\subsection{SHAP (SHapley Additive exPlanations) analysis for model interpretation}
Understanding how each parameter contributes to the model prediction is paramount for model deployment in real industrial applications. It determines the contribution score of each parameter, which is calculated by the Shapley values \cite{NIPS2017_8a20a862}. It calculates the impact of each parameter on the model prediction by considering all the parameter combinations and determining the marginal contribution of each parameter \cite{NIPS2017_8a20a862}. Shapley values are calculated by the Eq. \ref{eq: Shapley value}

\begin{equation} 
\label{eq: Shapley value}
\text{Shapley value } (\phi_j) = \sum_{S \subseteq N \setminus \{j\}} \frac{|S|!(|N| - |S| - 1)!}{|N|!} \left[g(S \cup \{j\}) - g(S)\right]
\end{equation}where: \( N \) is the set of all parameters, \( S \) is the subset of parameters excluding feature \(i\), and \( f(S) \) is the prediction by the model with parameters in the subset \( S \).

\section{Results and discussions}
This section analyzes the prediction of HTC on microchannel structured surfaces through an empirical approach, ML technique, and PIMLAF framework, compares their predictions, and evaluates the impact of different input parameters on the optimized model predictions. 

\subsection{Predictions by empirical correlations}

Empirical correlations are simple to use with low computation resources and can deliver quick results without complex simulations or iterations. Eighteen empirical correlations have been evaluated for predicting the boiling HTC on microchannel structured surfaces, and their performance across various metrics is compared in Table \ref{tab: Comparison_of_Correlations_for parallel microchannels}. The negative values of $R^2$ for some of the correlations indicate that the model is performing worse than the simple mean model of the data. All the properties used in the correlations are evaluated under saturated conditions. The formulations of these correlations are provided below, and the conditions under which these correlations are developed are outlined in Table \ref{tab: Empirical correlations and their operating conditions.}.

\begin{enumerate}[label=\arabic*.]

    \item Stephan and Abdelsalam correlation \cite{STEPHAN198073}:
\begin{flalign*}
& \begin{aligned}
&h_{\mathit{stephan, hydrocarbons}} = 0.0546 \cdot \frac{k_l}{D_d} \cdot \left( \frac{\rho_v}{\rho_l} \right)^{0.5} \cdot \left( \frac{qD_d}{k_l T_{sat}} \right)^{0.67} \cdot \left( \frac{\rho_l - \rho_v}{\rho_l} \right)^{-4.33} \cdot \left( \frac{h_{lv}D_d^2}{\alpha_l^2} \right)^{0.248} \\
&h_{\mathit{stephan, refrigerants}} = 207 \cdot \frac{k_l}{D_d} \cdot \left( \frac{qD_d}{k_l T_{sat}} \right)^{0.745} \cdot \left( \frac{\rho_v}{\rho_l} \right)^{0.581} \cdot \left( \frac{\nu_l}{\alpha_l} \right)^{0.53} \\
&h_{\mathit{stephan, other\ liquids}} = 0.23 \cdot \frac{k_l}{D_d} \cdot \left( \frac{qD_d}{k_l T_{sat}} \right)^{0.297} \cdot \left( \frac{h_{lv}D_d^2}{\alpha_l^2} \right)^{0.371} \cdot \left( \frac{\alpha_l^2 \rho_l}{\sigma D_d} \right)^{0.35} \cdot \left( \frac{\rho_l - \rho_v}{\rho_l} \right)^{-1.73} \\
\end{aligned} &
\end{flalign*}

    \item Jung correlation \cite{JUNG2003240}:
\begin{flalign*}
& \begin{aligned}
&h_{\mathit{jung}} = 10 \cdot \frac{k_l}{D_d} \cdot \left( \frac{(q/A)D_d}{k_l T_{sat}} \right)^{C_l} \cdot P_r^{0.1} \cdot (1 - T_r)^{-1.4} \cdot \left( \frac{\nu_l}{\alpha_l} \right)^{-0.25} \\
&\text{where} \\
&C_l = 0.855 \cdot \left( \frac{\rho_v}{\rho_l} \right)^{0.309} \cdot P_r^{-0.437}, \quad P_r = \frac{P_{op}}{P_c}, \quad T_r = \frac{T_{sat}}{T_c}
\end{aligned} &
\end{flalign*}

    \item Gorenflo and Kenning correlation \cite{gorenflo1993vdi}:
\begin{flalign*}
& \begin{aligned}
&h_{\mathit{gorenflo}} = {h_0} \cdot C \cdot F(P_r) \cdot \left( \frac{q}{q_0} \right)^n, \quad C = \left( \frac{R_a}{R_{a0}} \right)^{0.133}, \quad R_{a0} = 0.4 \, \mu \text{m}, \quad q_0 = 20,000 \, \text{W/m}^2, \\
&F(P_r) = 
\begin{cases} 
1.73 P_r^{0.27} + \left( 6.1 + \frac{0.68}{1 - P_r} \right) P_r^2 & \text{for water, with } n = 0.9 - 0.3P_r^{0.15}, \\
1.2 P_r^{0.27} + \left( 2.5 + \frac{1}{1 - P_r} \right) P_r & \text{for all other fluids, with } n = 0.9 - 0.3P_r^{0.3}.
\end{cases} \\
&h_0 \text{ is the reference heat transfer coefficient obtained from Gorenflo experimental data}
\\
& \text{for different fluids}
\end{aligned} &
\end{flalign*}

    \item Tarrad and Khudor correlation \cite{HussainTarrad2014ACF}:
\begin{flalign*}
& \begin{aligned}
&h_{\mathit{tarrad}} = 0.2411 \cdot h_0 \cdot \left( \frac{\rho_l h_{lv}^{3/2}}{q} \right)^{0.0864} \cdot \left( \frac{C_{pl} \cdot \sigma}{k_l h_{lv}^{0.5}} \right)^{1.40} \cdot \left( \frac{\rho_v}{\rho_l} \right)^{0.115} \cdot \lambda^{1.125} \cdot \left( \frac{P_{op}}{P_c} \right)^{-0.271} \\
&\text{where \( h_0 \) is the reference heat transfer coefficient as in the Gorenflo correlation.} \\
&\text{For fluids not covered by the Gorenflo correlation, \( h_0 \) is given by} \\
&h_0 = 0.1 \cdot P_c^{0.69} \cdot q^{0.7} \cdot F(P_r), \quad F(P_r) = 1.8 P_r^{0.17} + 4 P_r^{1.2} + 10 P_r^{10}, \text{where \( P_c \) in bar.}
\end{aligned} &
\end{flalign*}

    \item Shah correlation \cite{Balkrushna2022}:
\begin{flalign*}
& \begin{aligned}
&h_{\mathit{shah}} = 0.155 \cdot h_{\mathit{gorenflo}} \cdot \left( \frac{q_w}{\mu_l h_{lv}} \sqrt{\frac{\sigma}{(\rho_l - \rho_v) g}} \right)^{0.235} \cdot \left( \frac{P}{P_c} \right)^{-0.651} \cdot \\
&\quad \quad \quad \left( \frac{P_{op} \cdot D_d}{\mu_l \cdot h_{lv}^{1/2}} \right)^{-0.172} \cdot \left( \frac{\rho_v}{\rho_l} \right)^{-0.165} \cdot \lambda^{0.109} \\
\end{aligned} &
\end{flalign*}

    \item Stephan and Preusser correlation \cite{stephan1979warmeubergang}:
\begin{flalign*}
& \begin{aligned}
h_{\mathit{stephan-preusser}} &= 0.1 \cdot \left( \frac{k_l}{D_d} \right) \cdot \left( \frac{q D_d}{k_l T_{sat}} \right)^{0.67} \cdot \left( \frac{\rho_v}{\rho_l} \right)^{0.156} \cdot \left( \frac{h_{lv} D_d^2}{\alpha_l^2} \right)^{0.371} \cdot \left( \frac{\alpha_l^2 \rho_l}{\sigma D_d} \right)^{0.35} \cdot \left( \frac{\mu_l C_{pl}}{k_l} \right)^{-0.16}.
\end{aligned} &
\end{flalign*}

    \item Rosenhow correlation \cite{Rohsenow1952}:
\begin{flalign*}
& \begin{aligned}
&\triangle T_{\mathit{sat\_rosen}} = \left( \frac{h_{lv}}{C_{pl}} \right) \cdot C_{sf} \cdot \left( \left( \frac{q}{\mu_l \cdot h_{lv}} \right) \cdot L_{c} \right)^{0.33} \cdot \left( \mathit{Pr}_l \right)^{m+1}\\
&h_{\mathit{rosen}}= \frac{q}{\triangle T_{sat\_rosen}}
\end{aligned} &
\end{flalign*}

    \item Labuntsov correlation \cite{labuntsov1973heat}:
\begin{flalign*}
& \begin{aligned}
h_{\mathit{labuntsov}} = & 0.075 \left( 1 + \left( 10 \left( \frac{\rho_v}{\rho_l - \rho_v} \right)^{0.67} \right) \right) \quad \times \left( \left( \frac{\rho_l \cdot k_l^2}{\sigma \cdot \mu_l \cdot T_{sat}} \right)^{0.33} \right) \left( q \right)^{0.67}
\end{aligned} &
\end{flalign*}

    \item Kruzhilin correlation \cite{kruzhilin1947free}:
\begin{flalign*}
& \begin{aligned}
h_{\mathit{kruzhilin}} = \left( 0.082 \cdot \frac {k_l} {L_c} \right) \left( \left( \frac{h_{lv} \cdot q}{g \cdot T_{sat} \cdot k_l} \cdot \frac{\rho_v}{\rho_l - \rho_v} \right)^{0.7} \right) \left( \frac{T_{sat} \cdot C_{pl} \cdot \sigma \cdot \rho_l}{h_{lv}^2 \cdot \rho_v^2 \cdot L_{c}} \right)^{0.33} \left( \mathit{Pr}_l^{-0.45} \right)
\end{aligned} &
\end{flalign*}

    \item Kichigin and Tobilevich correlation \cite{kichigin1955generalization}:
\begin{flalign*}
& \begin{aligned}
h_{\mathit{kichi\_tobil}} = \left( \frac{k_l}{L_c} \right) \left( 3.25 \times 10^{-4} \right) \left( \mathit{Re} \right)^{0.6} \left( \mathit{Pr}_l \right)^{0.6} \left( \left( \frac{g \cdot L_c^3}{\nu_l^2} \right)^{0.125} \right) \cdot \left( \frac{P_{op}}{\left( g \cdot \sigma \cdot (\rho_l - \rho_v) \right)^{0.5}} \right)^{0.7}
\end{aligned} &
\end{flalign*}


    \item Borishansky correlation \cite{borishanskii1969correlation}:
\begin{flalign*}
& \begin{aligned}
&A^* = 0.1011 \cdot \left( P_c^{0.69} \right)\\
&F = 1.8 \cdot \left( P_{\mathit{r}}^{0.17} \right) + 4 \cdot \left( P_{\mathit{r}}^{1.2} \right) + 10 \cdot \left( P_{\mathit{r}}^{10} \right)\\
&h_{\mathit{borishansky}} = \left( A^* \right)^{3.33} \cdot \left( \triangle T \right)^{2.33} \cdot \left( F\right)^{3.33}
\end{aligned} &
\end{flalign*}

    \item Kutateladze and Borishanski correlation \cite{kutateladze1966concise}:
\begin{flalign*}
& \begin{aligned}
h_{\mathit{kuta\_boris}} = \left( 0.44 \cdot \frac{k_l}{L_c} \right) \left( \left( \frac{1 \times 10^{-4} \cdot q \cdot P_{op}}{g \cdot h_{lv} \cdot \rho_v \cdot \mu_l} \cdot \frac{\rho_l}{\rho_l - \rho_v} \right)^{0.7} \right) \left( \mathit{Pr}_l^{0.35} \right)
\end{aligned} &
\end{flalign*}

    \item Modified Kutateladze correlation \cite{kutateladze1990heat}:
\begin{flalign*}
& \begin{aligned}
& h_{\mathit{modif\_kuta}} = \left( 3.37 \times 10^{-9} \cdot \frac{k_l}{L_c} \cdot \left( \frac{h_{lv}}{C_{pl} \cdot q} \right)^{-2} \cdot M^{*-1} \right)^{\frac{1}{3}}\\
& M^* = \frac{g \cdot \sigma}{(\rho_l - \rho_v) \cdot \left( \frac{P_{op}}{\rho_v} \right)^2} \\
\end{aligned} &
\end{flalign*}

    \item Pioro correlation \cite{Pioro1999}:
\begin{flalign*}
& \begin{aligned}
h_{\mathit{pioro}} = C_{s} \cdot \frac{k_l}{L_c} \cdot \left( \frac{q}{{h_{lv} \cdot \sqrt{\rho_v}} \cdot {\left( \sigma \cdot g \cdot (\rho_l - \rho_v) \right)^{0.25}}} \right)^{\frac{2}{3}} \cdot \mathit{Pr}_l^{n}
\end{aligned} &
\end{flalign*}

    \item Cooper correlation \cite{COOPER1984157}:
\begin{flalign*}
& \begin{aligned}
h_{\mathit{cooper}} = 55 \cdot \left( \mathit{P}_{\mathit{r}}^{0.12 - (0.2 \cdot \log_{10}(\mathit{R}_{\mathit{q}}))} \right) \cdot \left (- \log_{10}(\mathit{P}_{\mathit{r}})\right)^{-0.55} \cdot \left( \mathit{M}^{-0.5} \right) \cdot \left( \mathit{q}^{0.67} \right)
\end{aligned} &
\end{flalign*}

    \item Cornwell–Houston correlation \cite{Cornwell1994}:
\begin{flalign*}
& \begin{aligned}
&h_{\mathit{cornwell}} = 9.7 \cdot \frac{k_l}{L_c} \cdot F_p \cdot \mathit{P}_{\mathit{c}}^{0.5} \cdot (Re)^{0.67} \cdot (\mathit{Pr}_l)^{0.4} \\
&F_p = 1.8 \cdot \mathit{P}_{\mathit{r}}^{0.17} + 4 \cdot \mathit{P}_{\mathit{r}}^{1.2} + 10 \cdot \mathit{P}_{\mathit{r}}^{10}
\end{aligned} &
\end{flalign*}

    \item Ribatski and Jabardo correlation \cite{Ribatski2003}:
\begin{flalign*}
& \begin{aligned}
&h_{\mathit{ribatski}} = 100 \cdot (\mathit{q}^{m}) \cdot (\mathit{P}_{\mathit{r}}^{0.45}) \cdot (-\log(\mathit{P}_{\mathit{r}}))^{-0.8} \cdot (\mathit{R}_{\mathit{q}}^{0.2}) \cdot (\mathit{M}^{-0.5}) \\
&m = 0.9 -  0.3 \cdot (\mathit{P}_{\mathit{r}}^{0.2}) 
\end{aligned} &
\end{flalign*}

\end{enumerate}

\begin{longtable}{lp{10cm}}
\caption{Empirical correlations and the parameters used for the development of correlations.}
\label{tab: Empirical correlations and their operating conditions.} \\
\toprule
\textbf{Authors} & \textbf{Parameters used for development of correlations} \\
\midrule
\endfirsthead

\toprule
\textbf{Authors} & \textbf{Parameters used for development of correlations} \\
\midrule
\endhead

\midrule
\multicolumn{2}{r}{\textit{Continued on next page}} \\
\midrule
\endfoot

\bottomrule
\endlastfoot

Rosenhow \cite{Rohsenow1952} & Fluids: Water, R-11, R-12, R-113, propane, n-pentane, ethanol, iso-Propanol, n-Butanol, 30\%, and 50\% potassium carbonate, carbon tetrachloride, benzene, n-Heptane, acetone.\newline Substrates: Copper, aluminum, platinum wires, zinc, nickel, inconel, stainless steel, brass, chromium \\
\midrule

Gorenflo and Kenning \cite{gorenflo1993vdi} &Fluids: Halogenated refrigerants \newline Substrates: Horizontal cylindrical surfaces.\\
\midrule

Stephan and Abdelsalam \cite{STEPHAN198073} & 5000 data points. \\ &Fluids: Hydrocarbons, water, refrigerants, and cryogenic fluids. \newline Substrates: Horizontal flat plate, cylinder, wire, tube - Copper, brass, platinum, nickel, stainless steel, chromium plated copper, inconel, nickel plated copper, bronze, german silver, gold coated ARMCO-iron.\\
\midrule

Jung \cite{JUNG2003240} & Fluids: HFC134a, CFC11, HFC32, HCFC123, HCFC22, HCFC142b, CFC12, and HFC125. \newline Substrates: Smooth horizontal copper tube.\\
\midrule

Tarrad and Khudor \cite{HussainTarrad2014ACF} & Fluids: Water, n-pentane, ethanol, R-123, R134a, R-113, R-114, R124, R-11, R-12, R-22. \newline Substrates: Low-finned tubes\\
\midrule

Shah \cite{Balkrushna2022} & Fluids: Water, R-123, R-141b, R-134a.\newline Substrates: Micro-finned cylindrical surfaces.\\
\midrule

Stephan and Preusser \cite{stephan1979warmeubergang} & Fluids: Water, organic fluids, binary and ternary mixtures \newline Substrates: Horizontal tubes\\
\midrule

Labuntsov \cite{labuntsov1973heat} & Can be used for a different set of fluids. \\
\midrule

Kichigin and Tobilevich \cite{kichigin1955generalization} & Fluids: Water and concentrated solutions. \newline Substrates: Steel tubes. \\
\midrule

Kruzhilin \cite{kruzhilin1947free}  & Fluids: Water and refrigerants. \newline Substrates: horizontal flat plates of different materials. \\
\midrule

Borishansky \cite{borishanskii1969correlation} & Fluids: Water, ethanol, and other fluids. \newline Substrates: Horizontal flat plates and tubes. \\
\midrule

Kutateladze and Borishanski \cite{kutateladze1966concise} & Can be used for a different set of fluids and large heat flux conditions. \\
\midrule

Kutateladze \cite{kutateladze1990heat} & Can be used for a different set of fluids. \\
\midrule

Pioro \cite{Pioro1999} & Fluids: Water, ethanol, iso-propanol, Propane, n-pentane, Benzene, n-butanol, R-113, R-11, R-12, carbon tetrachloride, n-heptane, 30\%, and 50\% Potassium carbonate, acetone \newline Substrates: Copper, aluminum, brass, chromium, platinum wires, stainless steel, zinc, nickel, inconel. Modified fluid surface parameter of Rohsenhow Correlation. \\
\midrule

Ribatski and Jabardo \cite{Ribatski2003} & 2600 data points. \newline Fluids: R-22, R-11, R-134a, R-11, R-12, R-123. \newline Substrates: Cylindrical surfaces - Brass, stainless steel, and copper.\\
\midrule

Cooper \cite{COOPER1984157} & 5641 data points. \newline Fluids: Water, R-113, R-12, R-114, ethanol, benzene, propane, water, cryogens - Hydrogen, neon, nitrogen, helium, oxygen. \newline Substrates: Aluminum, brass, copper, stainless steel, sodium-potassium alloy, platinum wires, nickel. \\
\midrule

Cornwell–Houston \cite{Cornwell1994} & Fluids: R-11, R-12, R-22, R-113, R-114, R-115, R-22, pentane, nonane, propane, ethane, hexane, benzene, ethanol, methanol, isobutanol, p-xylene, water. \newline Substrates: Horizontal tubes and tube bundles of various materials. \\
\midrule

\end{longtable}

From Table \ref{tab: Comparison_of_Correlations_for parallel microchannels}, we can conclude that correlation by Stephan \& Abdelsalam, Gorenflo, Labutsnov, Stephan \& Preusser, and modified correlation by Kutateladze exhibits $R^2$ value ranging from 0.3 to 0.5. This reduced performance may be due to its applicability to specific conditions.

\subsubsection{Proposed correlation for microchannel structured surfaces:}
\begin{figure}[H]
\centering
\subfloat[]{{\includegraphics[width=9cm]{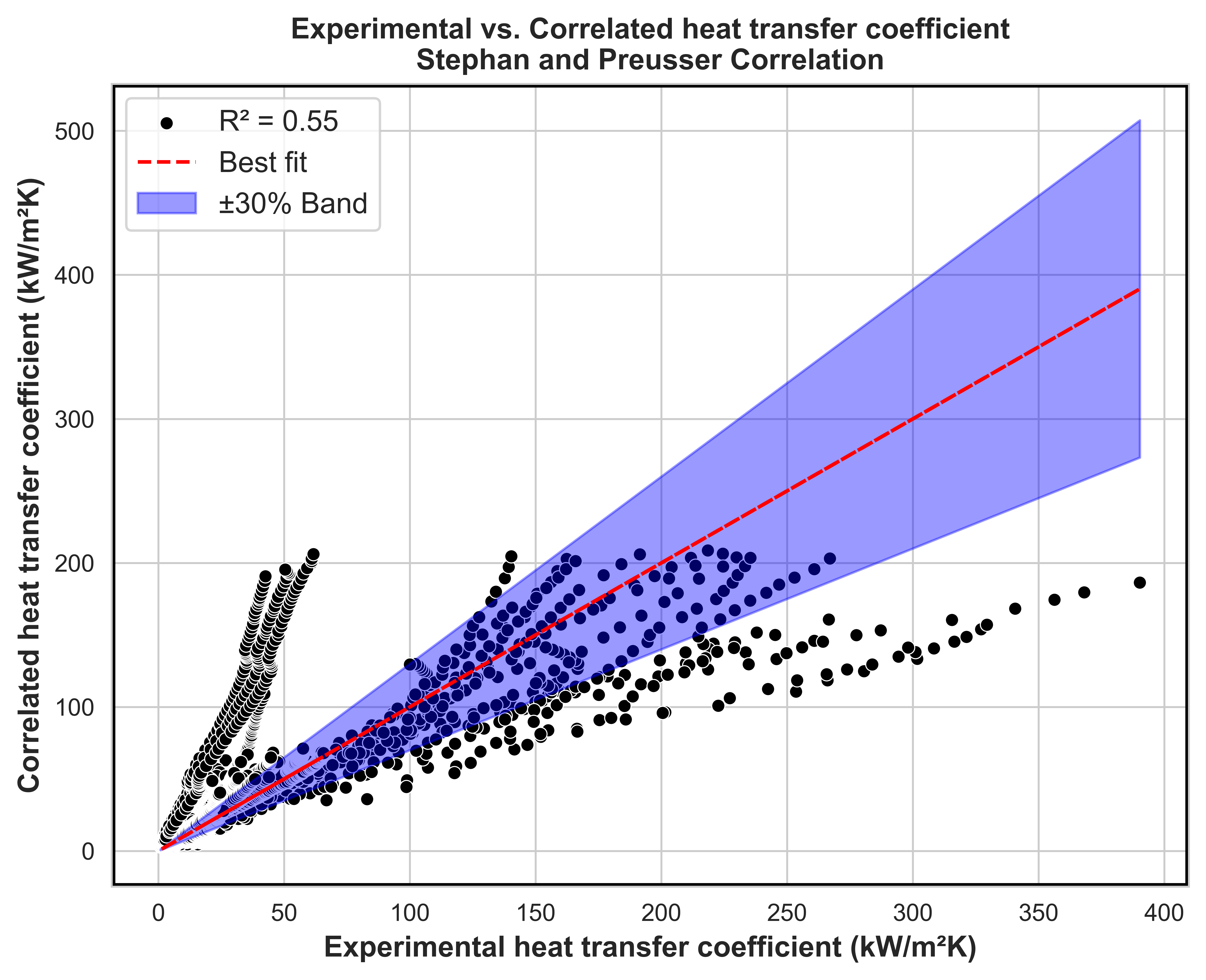}}\label{fig: stephan_preusser_correlation}}
\subfloat[]{{\includegraphics[width=9cm]{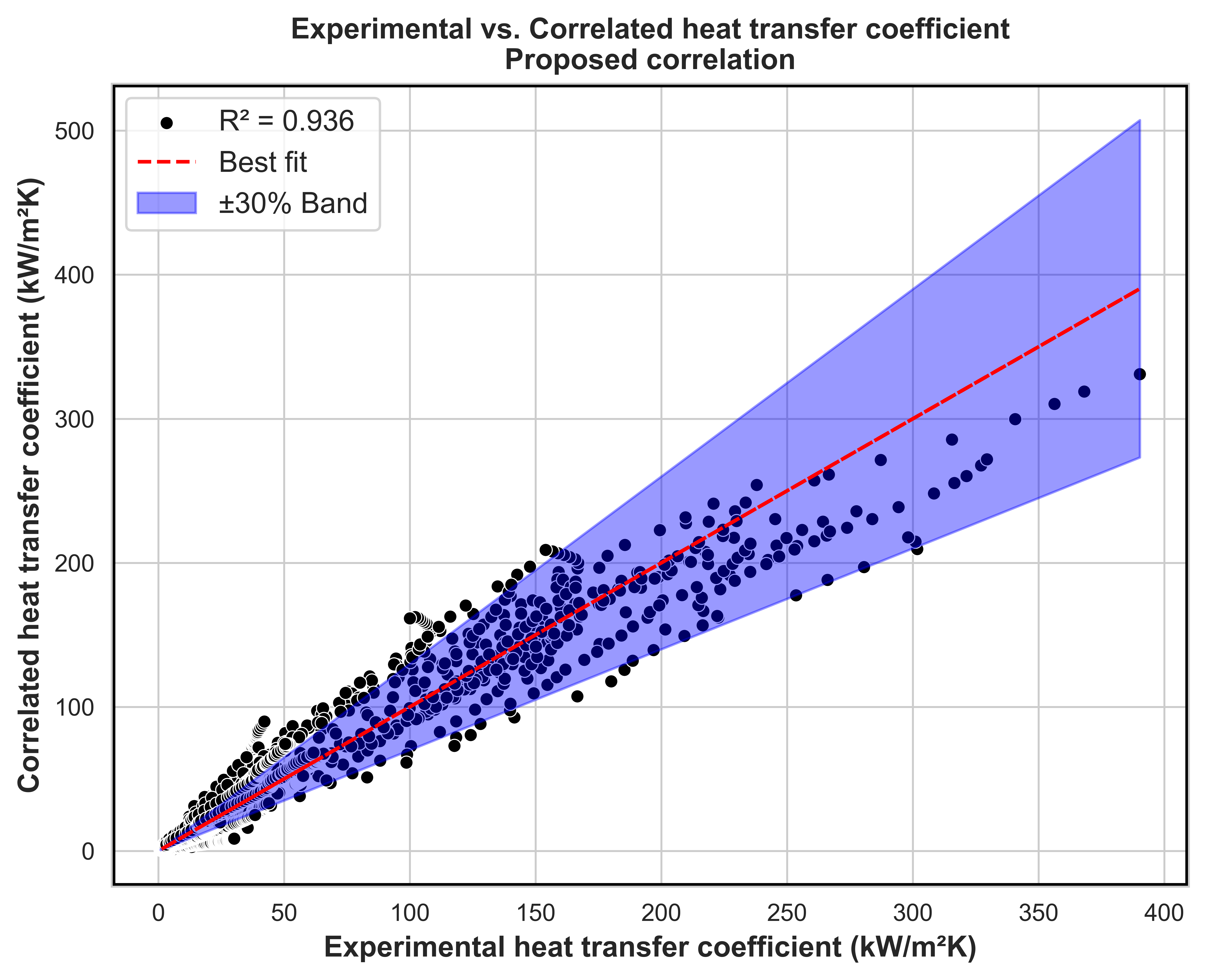}}\label{fig: proposed_microchannels_correlation}}\\
\caption{Performance of (a) Stephan-Preusser correlation and (b) proposed correlation (modified Stephan-Preusser correlation) for microchannel structured surfaces.}
\end{figure}

The lower predictive performance of the above correlations on the microchannel dataset may have resulted from the absence of incorporation of various influential parameters. To bridge this gap, nine dimensionless parameters ($\lambda$, $\frac{k_w}{k_l}$, $\frac{R_q}{r_{\mathit{cav}}}$, $\frac{\theta}{90}$, $P_{\mathit{r}}$, $\frac{M_f}{M_w}$, $\frac{h_f}{w_f}$, $\frac{w_g}{\mathit{p}}$, $\frac{D_h}{\mathit{p}}$)  have been defined to develop a new correlation. Among the above correlations, Stephan and Preusser's correlation displayed the highest $R^2$ value of 0.55. Thus, this empirical correlation has been modified with the addition of the above dimensionless parameters, and appropriate coefficients have been determined. The proposed correlation in Eq.(\ref{eq: Proposed Correlation for parallel microchannels}) is able to predict the microchannel structured surfaces dataset with a $R^2$ value of 0.936 and MAE of 4.94. Fig. \ref{fig: stephan_preusser_correlation} and \ref{fig: proposed_microchannels_correlation} show the improved performance of the proposed correlation over the Stephan-Preusser correlation. These metrics are evaluated based on the entire dataset of microchannel structured surfaces. 

The proposed correlation is given by Eq. \ref{eq: Proposed Correlation for parallel microchannels}:

\begin{equation}
\label{eq: Proposed Correlation for parallel microchannels}
\begin{aligned}
h_{\mathit{proposed}} = & \lambda^{0.472} \cdot \left( \frac{k_w}{k_l} \right)^{0.966} \cdot \left( \frac{R_q}{r_{\mathit{cav}}} \right)^{-0.197} \cdot \left( \frac{\theta}{90} \right)^{0.138} \cdot (P_{\mathit{r}})^{1.106} \\
& \cdot \left( \frac{M_f}{M_w} \right)^{-2.175} \cdot \left( \frac{h_f}{w_f} \right)^{-0.484} \cdot \left( \frac{w_g}{\mathit{p}} \right)^{0.295} \cdot \left( \frac{D_h}{\mathit{p}} \right)^{0.833} \\
& \cdot \left( 0.1 \cdot \frac{k_l}{D_d} \right) \cdot \left( \frac{q \cdot D_d}{k_l \cdot T_{sat}} \right)^{0.67} \cdot \left( \frac{\rho_v}{\rho_l} \right)^{0.156} \\
& \cdot \left( \frac{h_{lv} \cdot D_d^2}{\alpha_l^2} \right)^{0.371} \cdot \left( \frac{\alpha_l^2 \cdot \rho_l}{\sigma \cdot D_d} \right)^{0.35} \cdot \left( \frac{\mu_l \cdot C_{pl}}{k_l} \right)^{-0.16}
\end{aligned} 
\end{equation}

\begin{table}[H]
    \centering
    \caption{Comparison of correlations for microchannel structured surfaces.}
    \label{tab: Comparison_of_Correlations_for parallel microchannels}
    \begin{adjustbox}{max width=\textwidth}
    \renewcommand{\arraystretch}{1.5} 
    \begin{tabular}{llllllll}
        \hline
        \multirow{2}{*}{\textbf{Correlations}} & \multirow{2}{*}{\textbf{R\textsuperscript{2}}} & \multirow{2}{*}{\textbf{MAE}} & \multirow{2}{*}{\textbf{RMSE}} & \multicolumn{4}{c}{\textbf{\% Data Points within}} \\ \cline{5-8}
        & & & & \textbf{±10\%} & \textbf{±20\%} & \textbf{±30\%} & \textbf{±40\%} \\ \hline
        Rosenhow \cite{Rohsenow1952} & -32809514.6 & 167531.5 & 203845.8 & 0.0 & 0.0 & 0.0 & 0.0 \\ 
        Stephan-Abdelsalam \cite{STEPHAN198073} & 0.38 & 15.75 & 28.06 & 12.64 & 24.13 & 34.53 & 43.65 \\ 
        Jung \cite{JUNG2003240} & -0.44 & 22.96 & 42.63 & 11.67 & 23.10 & 34.54 & 43.83 \\ 
        Gorenflo and Kenning \cite{gorenflo1993vdi} & 0.38 & 11.18 & 28.06 & 10.87 & 21.86 & 38.88 & 48.81 \\ 
        Tarrad and Khudor \cite{HussainTarrad2014ACF} & 0.11 & 15.29 & 33.67 & 6.58 & 14.73 & 21.37 & 26.60 \\ 
        Shah \cite{Balkrushna2022} & -958.2 & 585.1 & 1102.2 & 0.0 & 0.0 & 0.0 & 0.0 \\ 
        Stephan-Preusser \cite{stephan1979warmeubergang} & 0.55 & 11.50 & 23.95 & 14.11 & 28.20 & 42.89 & 54.70 \\ 
        Labuntsov \cite{labuntsov1973heat} & 0.41 & 12.62 & 27.31 & 5.50 & 11.01 & 16.91 & 28.48 \\ 
        Kichigin and Tobilevich \cite{kichigin1955generalization} & -0.41 & 23.04 & 42.24 & 0.0 & 0.0 & 0.0 & 0.0 \\ 
        Kruzhilin \cite{kruzhilin1947free} & -338.41 & 362.91 & 655.64 & 0.0 & 0.0 & 0.0 & 0.0 \\ 
        Borishansky \cite{borishanskii1969correlation} & -0.87 & 24.76 & 48.63 & 0.60 & 1.33 & 2.12 & 3.21 \\ 
        Kutateladze and Borishanski \cite{kutateladze1966concise} & -0.36 & 22.56 & 41.45 & 0.0 & 0.0 & 0.0 & 0.0 \\ 
        Modified Kutateladze \cite{kutateladze1990heat} & 0.44 & 11.84 & 26.65 & 6.51 & 15.64 & 23.02 & 30.22 \\ 
        Pioro \cite{Pioro1999} & -1334090.4 & 17813.2 & 41105.0 & 0.0 & 0.0 & 0.0 & 0.0 \\ 
        Ribatski and Jabardo \cite{Ribatski2003} & 0.14 & 15.18 & 32.96 & 1.32 & 4.69 & 12.85 & 24.02 \\ 
        Cooper \cite{COOPER1984157} & 0.16 & 13.22 & 32.58 & 10.76 & 19.92 & 28.54 & 35.03 \\ 
        Cornwell-Houston \cite{Cornwell1994} & -48.62 & 110.17 & 250.69 & 3.93 & 8.04 & 12.16 & 15.80 \\  
        Proposed correlation & 0.936 & 4.94 & 8.98 & 22.19 & 44.81 & 68.50 & 78.70 \\ 
        \hline
    \end{tabular}
    \end{adjustbox}
\end{table}

\subsection{Predictions by Machine learning alogrithms}
Initially, the dataset is divided into training and testing datasets, with 80\% of the data for model training and the remaining 20\% for model testing. Various regression ML models have been trained on the microchannel structured surfaces dataset after data preprocessing using the Scikit-learn \cite{JMLR:v12:pedregosa11a} library. During training, these models learn the relationship among input variables by reducing the loss function. Table \ref{tab: Comparison_of_Regression_Models} shows the different models employed in this study. These models have been trained and then assessed based on the 5-fold cross-validation results. Although ET, KNN, XGBoost, RF, and LightGBM models rank at the top among these algorithms, their $R^2$ values on the test dataset are only around 0.91 after hyperparameter optimization through random search and 5-fold cross-validation. Table \ref{tab: Comparison_of_Regression_Models} shows the performance of these nineteen algorithms.

\begin{figure}[H]
\centering
\subfloat[]{{\includegraphics[width=9cm]{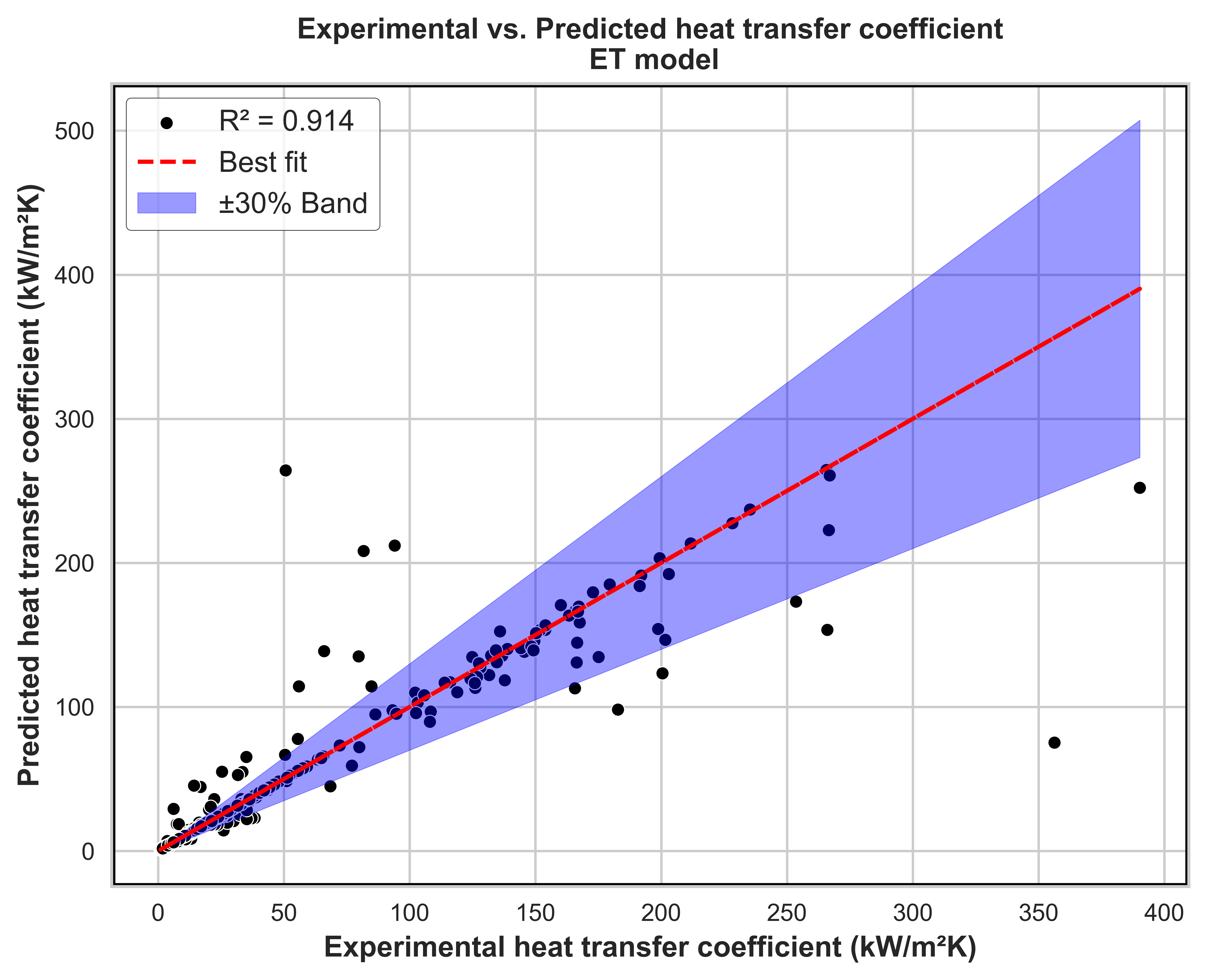}}\label{fig: microchannel_et_model_parity_plot}}
\subfloat[]{{\includegraphics[width=9cm]{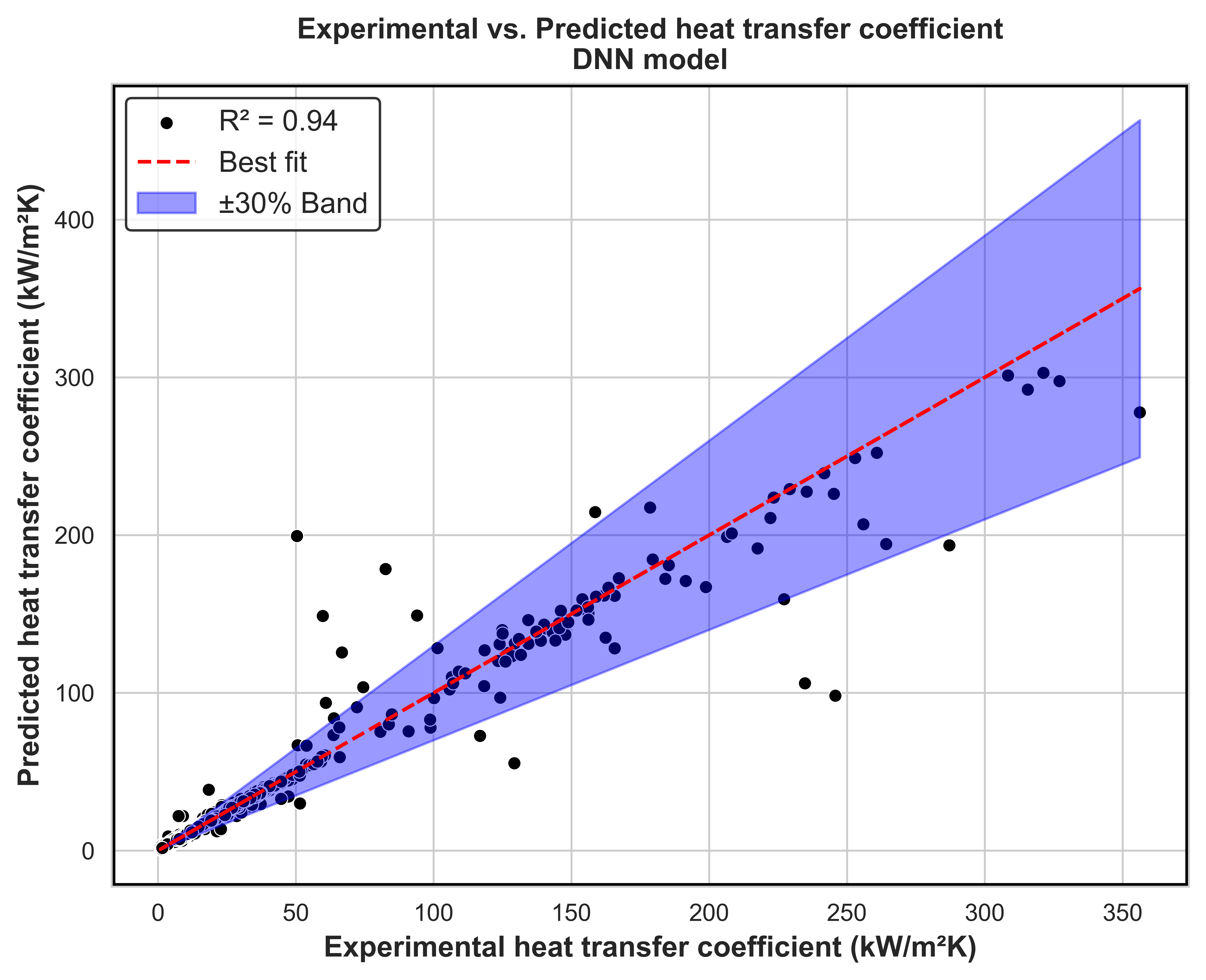}}\label{fig:microchannels_DNN_parity_plot}}\\
\caption{Performance of (a) Extra tree regression model and (b) DNN model for microchannel structured surfaces.}
\end{figure}

To improve the prediction accuracy, Deep neural network (DNN) models have been adopted. Deep neural networks can understand the non-linear patterns in the data with the help of activation functions and hidden layers. For this study, we adopted a 10-layer neural network architecture consisting of 120 neurons per layer and employed the Exponential Linear Unit (ELU) as the activation function for better gradient flow during backpropagation \cite{clevert2016fastaccuratedeepnetwork}. The model was optimized using the Adam optimizer \cite{kingma2017adammethodstochasticoptimization} with a learning rate of 0.001. To prevent overfitting, L1 and L2 regularization were applied with regularization coefficients of 0.001. The model was trained for 10,000 epochs to achieve optimal performance. These are the fine-tuned hyperparameters through random search optimization. By implementing the DNN model on the test dataset, the performance is increased with a $R^2$ value of 0.940. The DNN model is implemented using the Pytorch framework \cite{paszke2019pytorchimperativestylehighperformance}. Table \ref{tab: Comparison of DNN and top performing regression models.} shows the improved performance of the DNN model. Figs. \ref{fig: microchannel_et_model_parity_plot} and \ref{fig:microchannels_DNN_parity_plot} represent the parity plot of predicted and experimental HTC of microchannel structured surfaces dataset, corresponding to the ET regressor and DNN model, respectively.

\begin{table}[H]
    \centering
    \caption{Performance comparison of regression models.}
    \label{tab: Comparison_of_Regression_Models}
    \begin{adjustbox}{max width=\textwidth}
    \renewcommand{\arraystretch}{1.5} 
    \begin{tabular}{lccc}
        \hline
        \textbf{Model} & \textbf{R\textsuperscript{2}} & \textbf{MAE} & \textbf{RMSE} \\ \hline
        Extra trees regressor & 0.912 & 2.521 & 12.031 \\
        KNN regressor & 0.910 & 2.634 & 12.651 \\
        Extreme gradient boosting & 0.909 & 2.231 & 10.289 \\
        Random forest regressor & 0.903 & 2.029 & 10.732 \\
        Light gradient boosting machine & 0.901 & 3.102 & 10.942 \\
        Decision tree regressor & 0.871 & 2.030 & 11.853 \\
        Gradient boosting regressor & 0.822 & 5.705 & 14.823 \\
        CatBoost regressor & 0.817 & 2.391 & 9.461 \\
        AdaBoost regressor & 0.710 & 9.194 & 18.773 \\
        Linear regression & 0.616 & 10.800 & 21.791 \\
        Ridge regression & 0.607 & 11.233 & 22.034 \\
        Bayesian ridge & 0.607 & 11.239 & 22.041 \\
        Lasso regression & 0.541 & 12.025 & 23.852 \\
        Lasso least angle regression & 0.541 & 12.027 & 23.853 \\
        Huber regressor & 0.503 & 9.699 & 24.888 \\
        Elastic net & 0.464 & 12.605 & 25.813 \\
        Passive aggressive regressor & 0.458 & 14.316 & 25.784 \\
        Orthogonal matching pursuit & 0.353 & 14.159 & 28.296 \\
        Dummy regressor & -0.002 & 17.798 & 35.250 \\ \hline
    \end{tabular}
    \end{adjustbox}
\end{table}

\begin{table}[H]
    \centering
    \caption{Comparison of DNN and top performing regression models.}
    \label{tab: Comparison of DNN and top performing regression models.}
    \begin{adjustbox}{max width=\textwidth}
    \renewcommand{\arraystretch}{1.5} 
    \begin{tabular}{lccc}
        \hline
        \textbf{Model} & \textbf{R\textsuperscript{2}} & \textbf{MAE} & \textbf{RMSE} \\ \hline
        Deep neural network model & 0.940 & 2.037 & 9.913 \\
        Extra Trees regressor & 0.912 & 2.521 & 12.031 \\
        KNN regressor & 0.910 & 2.634 & 12.651 \\
        Extreme gradient boosting & 0.909 & 2.231 & 10.289 \\
        Random forest regressor & 0.903 & 2.029 & 10.732 \\
        Light gradient boosting machine & 0.901 & 3.102 & 10.942 \\ \hline
    \end{tabular}
    \end{adjustbox}
\end{table}

\subsection{Predictions by hybrid Machine learning framework}
Both empirical correlations and the standalone DNN model encounter lower prediction accuracy of HTC in microchannel structured surfaces. Even though empirical correlations can provide domain knowledge, they are limited by the inability to under complex relationships in the data. Also, DNN may suffer from generalizability and may perform well only within the trained parametric ranges. Meanwhile, it can learn non-linear relationships in the data. To synergize and leverage the advantage of both these approaches, the PIMLAF framework described in section \ref{sec: Methodology} is employed. 

\begin{table}[!h]
    \centering
    \caption{Optimized hyperparameters of the neural network model employed in PIMLAF.}
    \label{tab:Hyperparameters employed in the neural network model (PIMLAF).}
    \begin{adjustbox}{max width=\textwidth}
    \renewcommand{\arraystretch}{1.5} 
    \begin{tabular}{lll}
        \hline
        \textbf{Hyperparameters} & \textbf{Values} & \textbf{Description} \\ \hline
        No. of hidden layers & 8 & Total number of fully connected (dense) layers in the network. \\
        No. of neurons & [90, 90, 90, 90, 90, 90, 90, 90] & Number of neurons in each hidden layer. \\
        Activation function & ELU (Exponential Linear Unit) & Activation function applied after each layer, with $\alpha=1.0$. \\
        Learning rate & 0.001 & Step size for weight updates during training. \\
        Optimizer & Adam & Optimization algorithm for gradient-based updates. \\
        $\lambda_1$ & 0.001 &  L1 loss regularization parameter. \\
        $\lambda_2$ & 0.001 &  L2 loss regularization parameter. \\
        Number of epochs & 10,000 & Total number of training iterations over the dataset. \\ 
        \hline
    \end{tabular}
    \end{adjustbox}
\end{table}

\begin{figure}[!h]    
\centering
    \includegraphics[width=18cm]{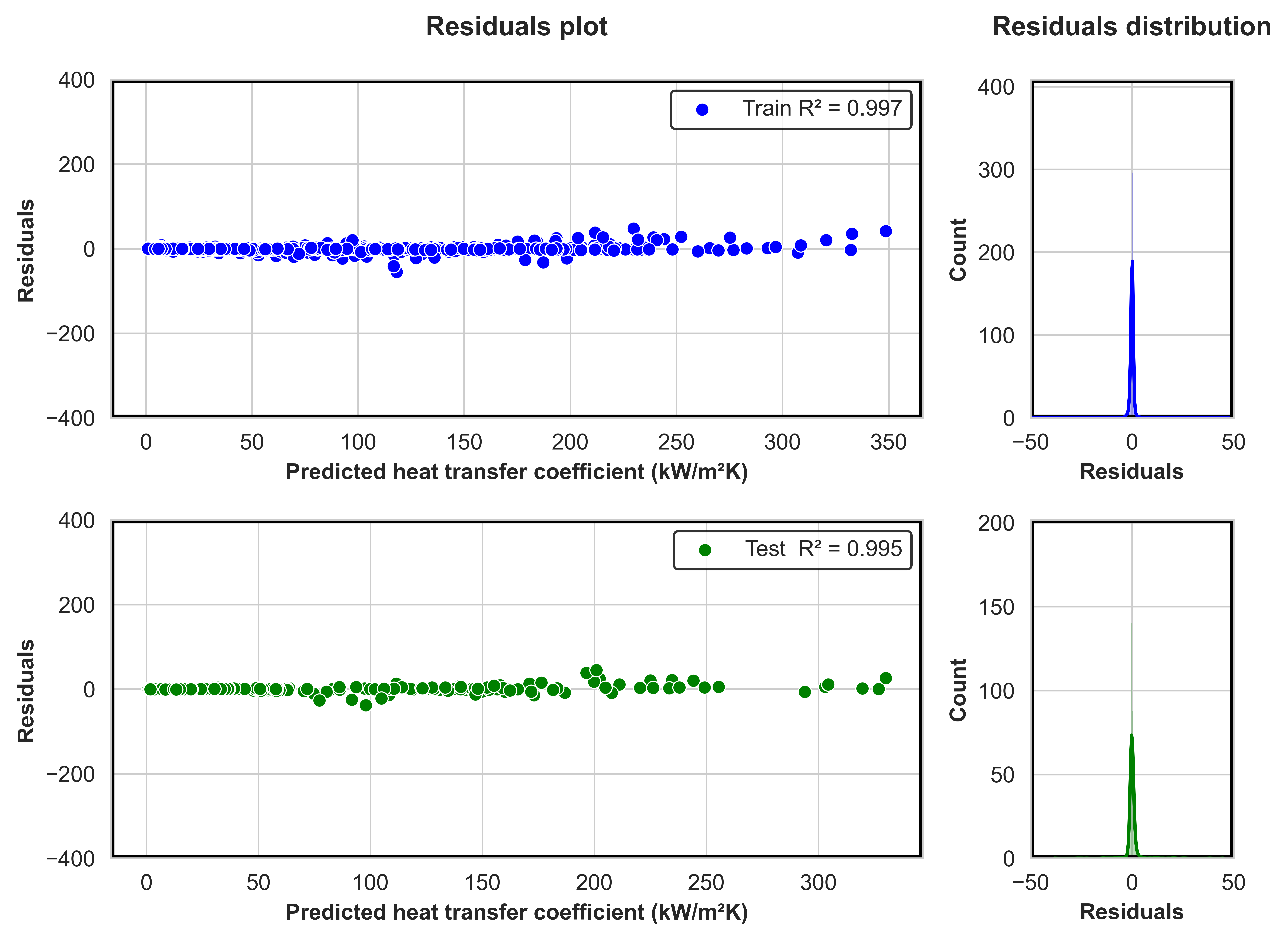}
\caption{Residual plot of PIMLAF model for microchannel structured surfaces data.}
\label{fig: residual_plot_PIMLAF}
\end{figure}

\begin{table}[!h]
    \centering
    \caption{Comparison of performance - Proposed correlation, machine learning models, and hybrid framework.}
    \label{tab: Performance comparison of proposed correlation, machine learning models, and hybrid framework.}
    \begin{adjustbox}{max width=\textwidth}
    \renewcommand{\arraystretch}{1.5} 
    \begin{tabular}{llllllll}
        \hline
        \multirow{2}{*}{\textbf{Models}} & \multirow{2}{*}{\textbf{R\textsuperscript{2}}} & \multirow{2}{*}{\textbf{MAE}} & \multirow{2}{*}{\textbf{RMSE}} & \multicolumn{4}{c}{\textbf{\% deviation of data within}} \\ \cline{5-8}
        & & & & \textbf{±10\%} & \textbf{±20\%} & \textbf{±30\%} & \textbf{±40\%} \\ \hline
        Extra Trees regressor & 0.912 & 2.521 & 12.031 & 78.752 & 90.182 & 94.250 & 96.774 \\ 
        Deep Neural Network (DNN) & 0.940 & 2.037 & 9.913 & 84.362 & 92.637 & 96.003 & 97.616 \\ 
        Hybrid Framework - Overall dataset & 0.995 & 0.907 & 2.999 & 88.640 & 96.283 & 97.756 & 98.808 \\ 
        Hybrid Framework - Water dataset & 0.992 & 2.403 & 6.347 & 89.655 & 97.318 & 98.084 & 98.084 \\ 
        Hybrid Framework - Other fluids dataset & 0.997 & 0.319 & 0.542 & 91.674 & 95.966 & 97.768 & 98.970 \\ 
        \hline
    \end{tabular}
    \end{adjustbox}
\end{table}

\begin{figure}[H]    
\centering
    \includegraphics[width=15cm, height = 13cm]{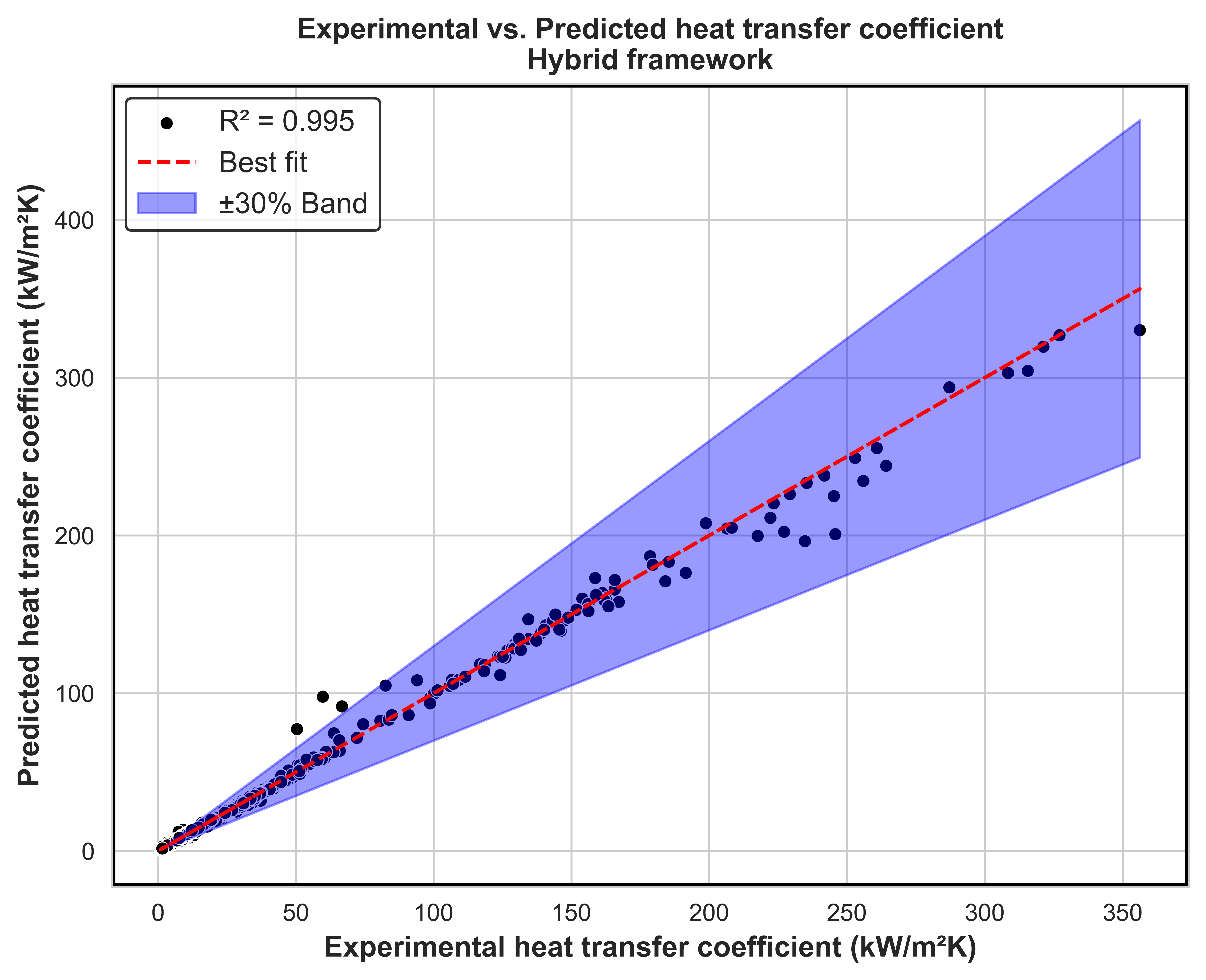}
\caption{Performance of PIMLAF model for microchannel structured surfaces.}
\label{fig: microchannel_PIMLAF_parity_plot}
\end{figure}

\begin{figure}[H]    
\centering
    \includegraphics[width=18cm]{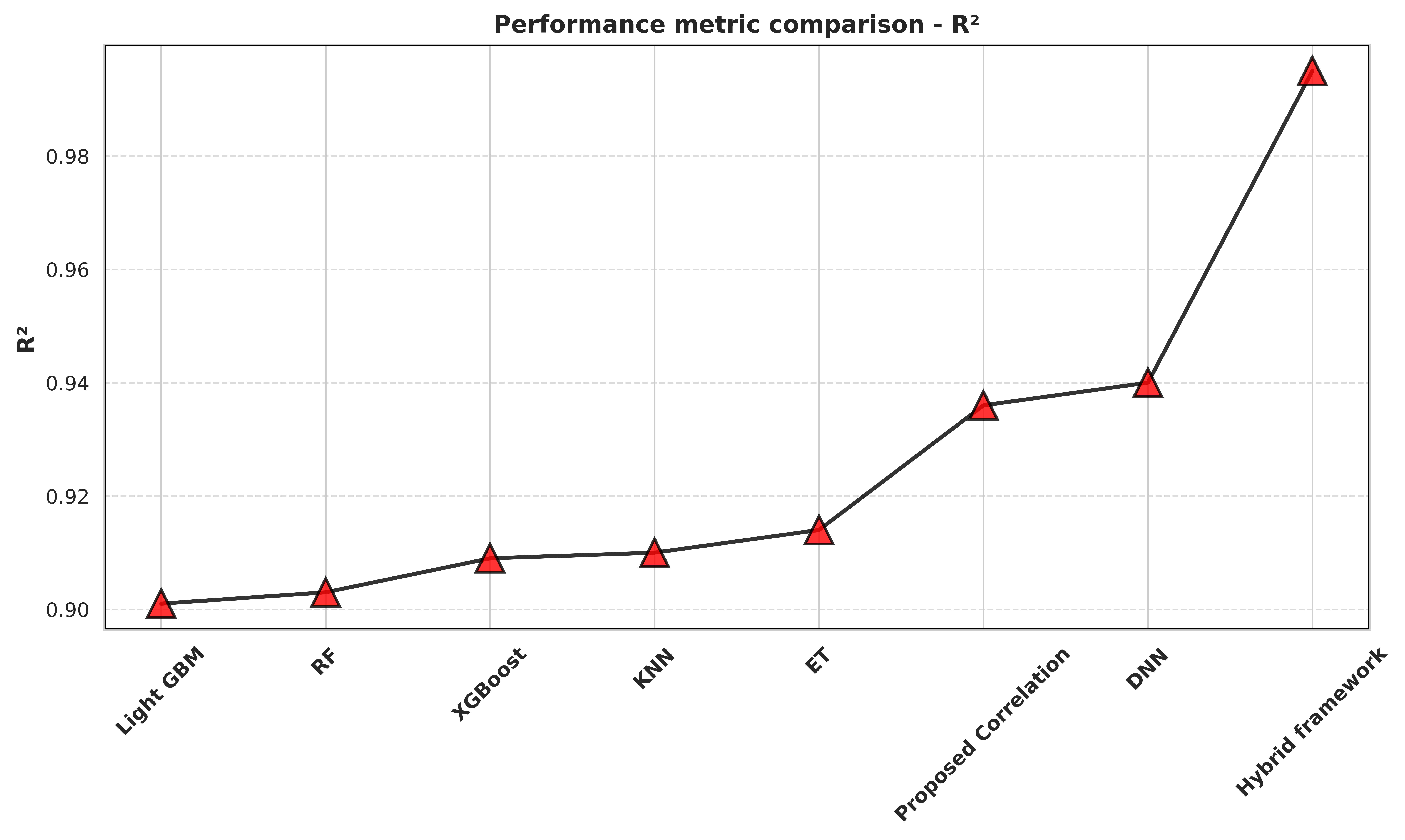 }
\caption{Performance metric comparison: Proposed correlation, ML model, and hybrid framework.}
\label{fig: performance_comparison_of_models}
\end{figure}

The new correlation proposed in this study is used to provide a good baseline prediction. Then, the DNN model with optimized hyperparameters as in Table \ref{tab:Hyperparameters employed in the neural network model (PIMLAF).} is developed to learn the correlation residual errors. Then, through backpropagation, the MSE loss function is minimized, yielding a highly accurate model for HTC prediction on microchannel structured surfaces with $R^2$, MAE, and RMSE values as 0.995, 0.907, and 2.999, respectively. Also, this model exhibits better performance with a $R^2$ value of around 0.99 when tested on a dataset that has undergone imputation for unknown feature values. Fig. \ref{fig: residual_plot_PIMLAF} illustrates the training and testing performance of this optimized physics-informed hybrid framework. The parity plot in Fig. \ref{fig: microchannel_PIMLAF_parity_plot} shows that there is better agreement between experimental and predicted HTC using the hybrid framework. Furthermore, this hybrid framework is implemented to predict HTC for the water and other fluids datasets, demonstrating better prediction with $R^2$ values of 0.992 and 0.997, respectively. Table \ref{tab: Performance comparison of proposed correlation, machine learning models, and hybrid framework.} presents the predictive performance comparison of the models adopted in this study for HTC estimation. All the model performances are based on the test dataset. Fig. \ref{fig: performance_comparison_of_models} indicates the superiority of the hybrid framework in comparison to other ML models and empirical correlation. 

This enhanced performance of the hybrid framework can be attributed to the effective harnessing of both domain-specific insights through proposed correlation and data-driven information machine learning techniques. Also, the comprehensive choice of input parameters for the ML model aids in capturing the complex relationships. With the prior domain knowledge in hand, this hybrid framework can generalize better to unseen datasets, making it more robust.

\subsection{Predictions for heat flux condition as input}

When heat flux is given as an input parameter instead of wall superheat, the proposed hybrid framework can be used through an iterative procedure. Starting with an initial guess for wall superheat, the HTC can be determined through the hybrid framework. With the guessed wall superheat and estimated HTC, heat flux can be calculated. When the calculated heat flux converges to the input heat flux, the iterations can be stopped. The final estimated HTC is the predicted value for the given input heat flux.

\subsection{SHAP interpretation of the optimized hybrid framework model}
SHAP interpretation explains the output of the predicted model by allocating a SHAP value to each feature in the model. This highlights the important features, feature interaction, and their impact on model prediction. SHAP summary plot, SHAP value plot, and SHAP dependency plot aid in understanding the feature significance. Positive and negative SHAP values indicate that the feature drives the prediction towards higher values and lower values, respectively. Zero SHAP value signifies no impact on model output. The SHAP summary plot (beeswarm plot) illustrates the global impact and variation of each feature, while the SHAP bar plot ranks the features based on their overall contribution based on the mean SHAP value of the corresponding feature. Red and blue values indicate the actual feature values, with red indicating higher values and blue indicating lower values. When the feature variation changes from blue on the left of the zero SHAP line to red on the right, it illustrates the direct impact on the prediction, and vice-versa otherwise. Table \ref{tab:Significant_Variables_Parallel_Microchannels} shows the major significant parameters affecting HTC when modeled with the overall data, water, and other fluids datasets.

\begin{table}[H]
\centering
\caption{Major parameters impacting heat transfer in microchannel structured surfaces.}
\label{tab:Significant_Variables_Parallel_Microchannels}
\begin{adjustbox}{max width=\textwidth}
\begin{tabular}{@{}lll@{}}
\toprule
\textbf{Overall dataset} & \textbf{Water dataset} & \textbf{Other fluids dataset} \\ \midrule
\hspace{35pt}$\triangle T$ & \hspace{35pt}$k_w$   & \hspace{35pt}$\triangle T$    \\
\hspace{35pt}$R_{q}$       & \hspace{35pt}$\theta$ & \hspace{35pt}$P_{film}$       \\
\hspace{35pt}$\theta$      & \hspace{35pt}$w_f$    & \hspace{35pt}$R_{q}$          \\
\hspace{35pt}$w_f$         & \hspace{35pt}$R_{q}$  & \hspace{35pt}$T_w$            \\
\hspace{35pt}$T_w$         & \hspace{35pt}$h_f$    & \hspace{35pt}$\lambda$            \\
\hspace{35pt}$\rho_l$      & \hspace{35pt}$p$      & \hspace{35pt}$w_g$        \\
\hspace{35pt}$k_{l}$       & \hspace{35pt}$w_g$    & \hspace{35pt}$h_f$            \\
\hspace{35pt}$P_{film}$    & \hspace{35pt}$k_{l}$  &  \hspace{35pt}$w_f$                              \\
\hspace{35pt}$h_f$         &                       &                                \\
\hspace{35pt}$w_g$         &                        &                                \\
\hspace{35pt}$\lambda$     &                        &                                \\ 
\hspace{35pt}$p$           &                        &                                \\ 
\bottomrule
\end{tabular}
\end{adjustbox}
\end{table}

\begin{figure}[H]    
\centering
    \includegraphics[width=11cm, height = 14cm]{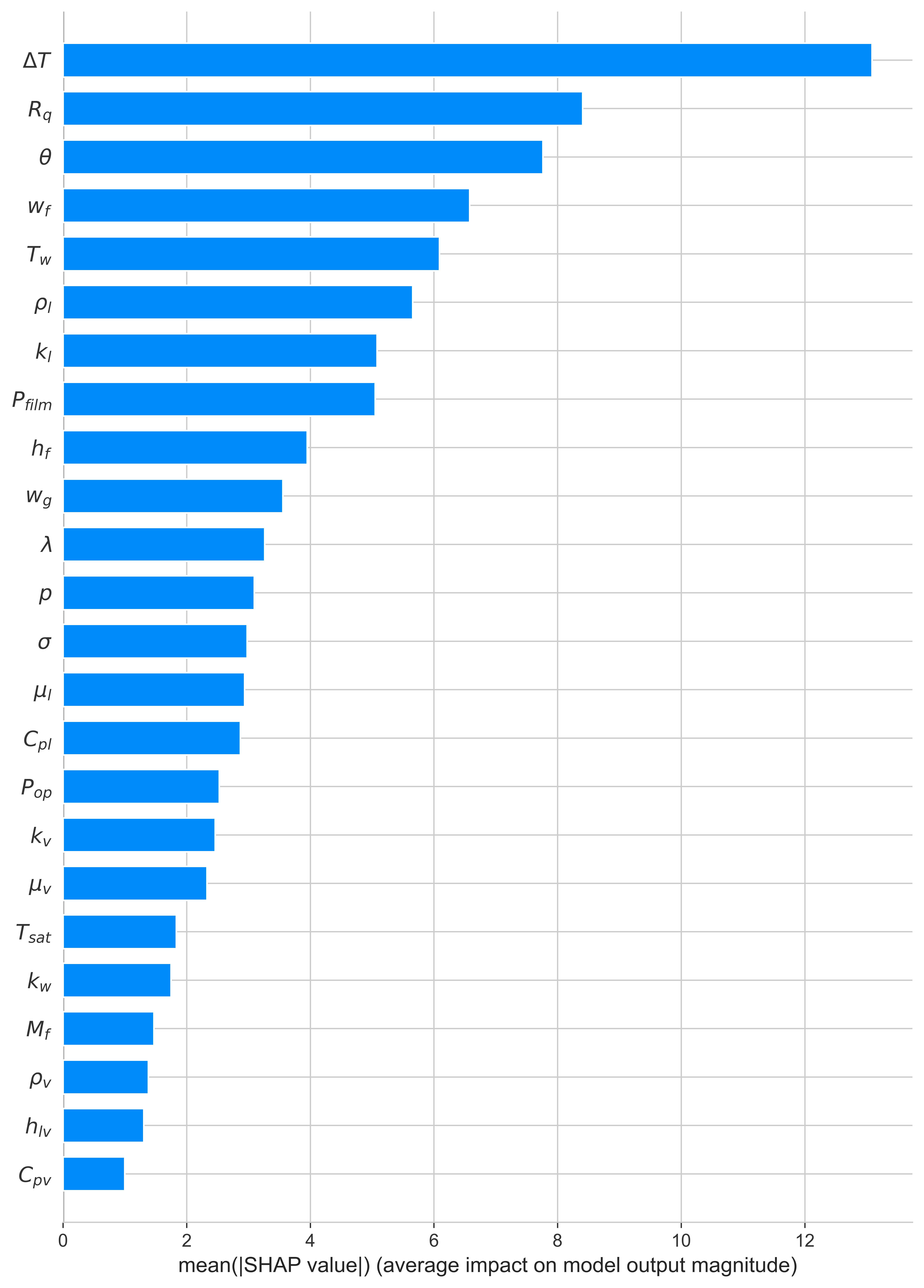}
\caption{SHAP summary plot of microchannel structured surfaces for the overall dataset - Bar plot.}
\label{fig: parallel_microchannels_bar_plot_overall_dataset}
\end{figure}

\begin{figure}[H]    
\centering
    \includegraphics[width=9cm, height = 12cm]{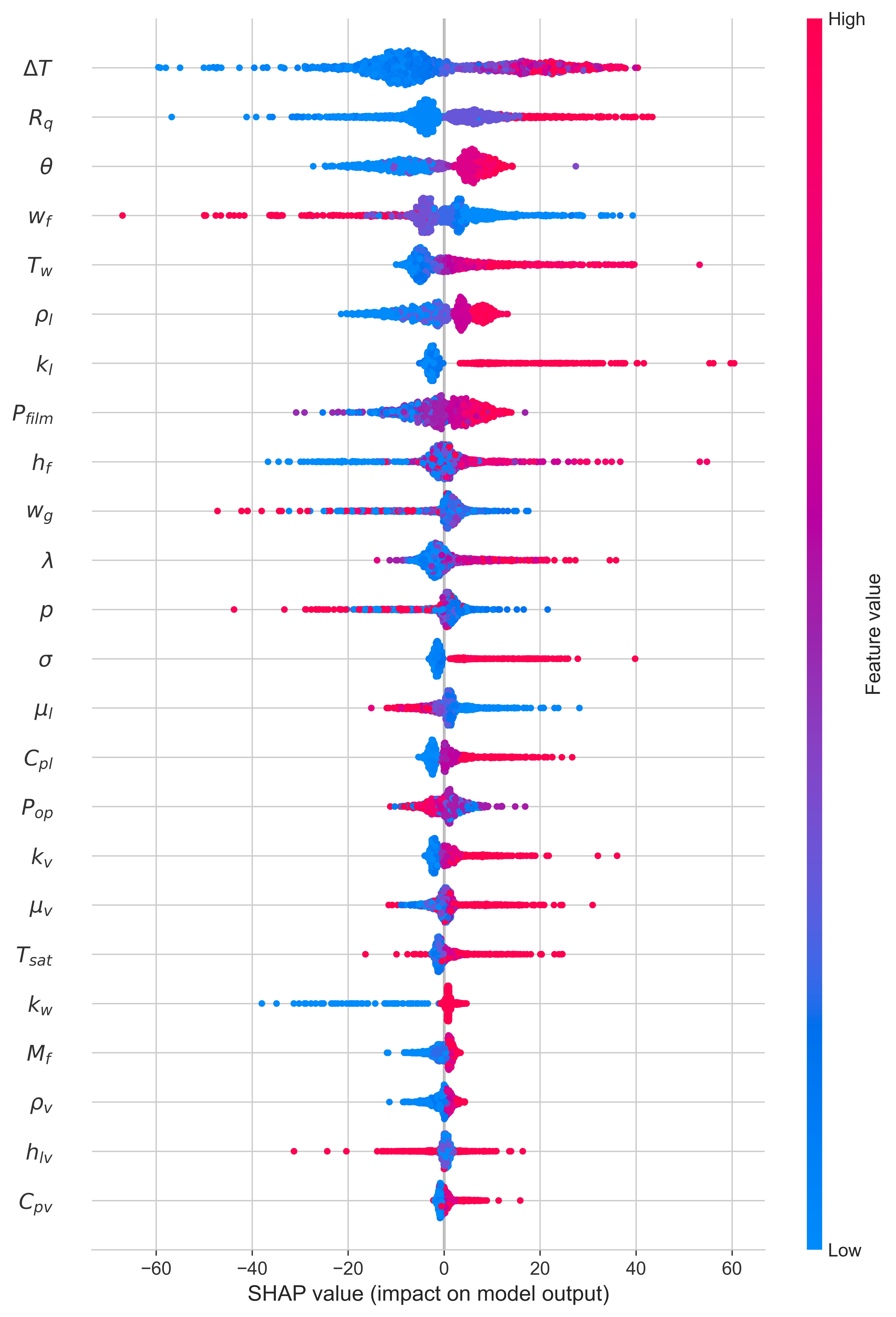}
\caption{SHAP summary plot of microchannel structured surfaces for the overall dataset - Beeswarm plot.}
\label{fig: parallel_microchannels_beeswarm_plot_overall_dataset}
\end{figure}

\begin{figure}[H]
\centering
\subfloat[]{{\includegraphics[width=9cm, height = 11cm]{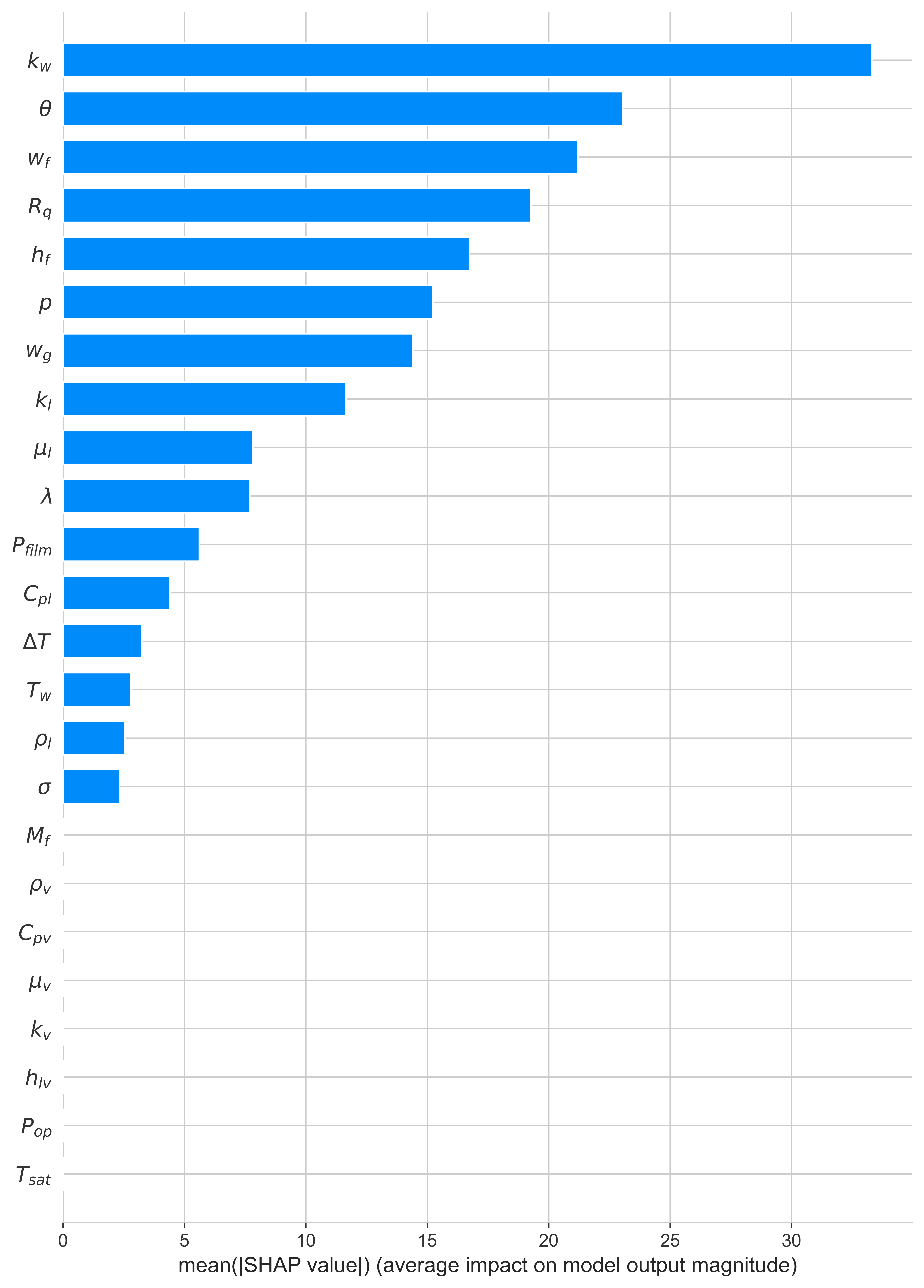}}\label{fig: parallel_microchannels_bar_plot_water}}
\subfloat[]{{\includegraphics[width=9cm, height = 11cm]{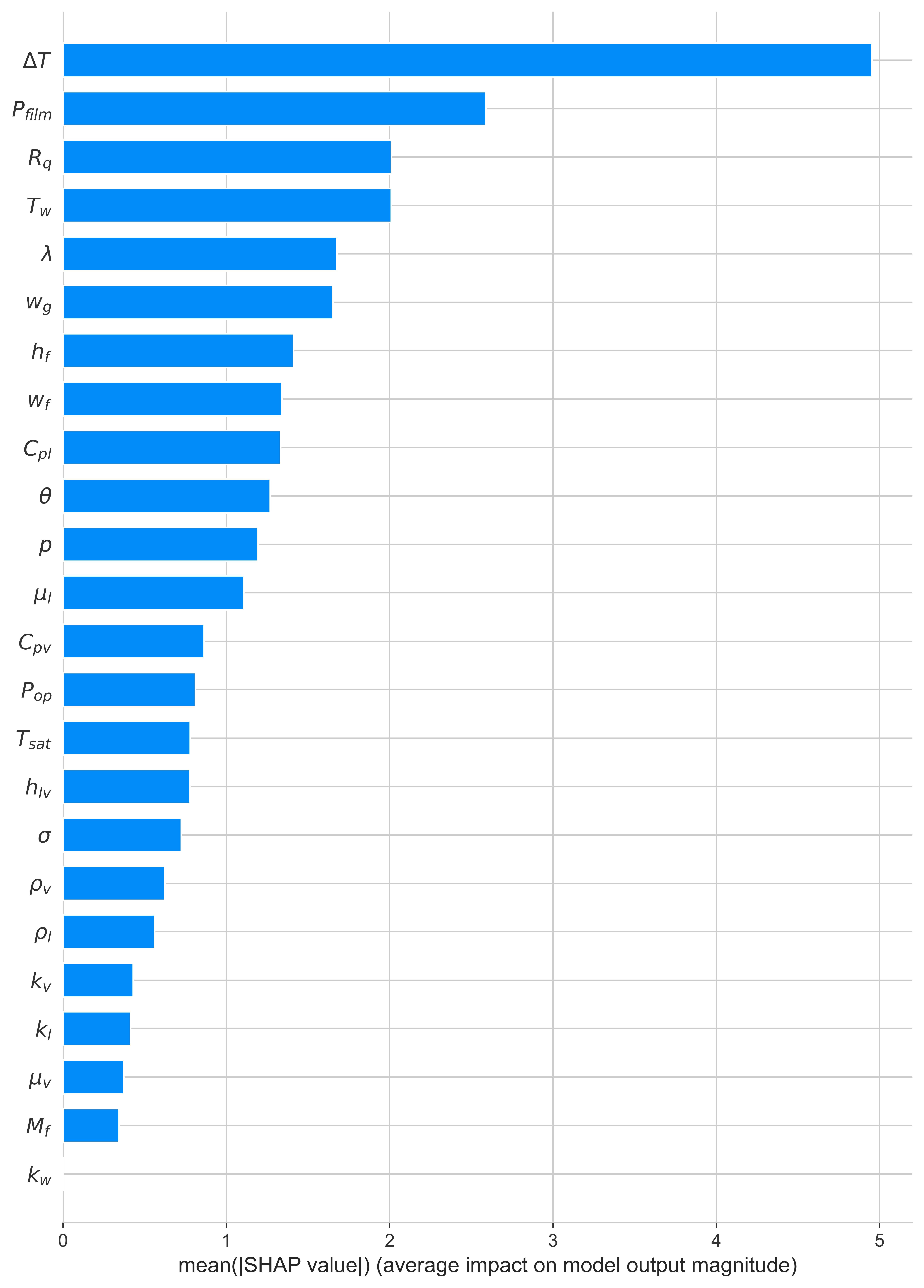}}\label{fig: parallel_microchannels_bar_plot_other_fluids}}
\caption{SHAP summary bar plot of microchannel structured surfaces for (a) water (b) other fluids.}
\end{figure}

\subsubsection{For the overall dataset} \label{subsubsection: overall_dataset}
The SHAP summary plot of the overall dataset (Figs. \ref{fig: parallel_microchannels_bar_plot_overall_dataset} and \ref{fig: parallel_microchannels_beeswarm_plot_overall_dataset}) highlights the following observations:

\begin{itemize}
    \item $\triangle T$ and $T_w$ exhibit a positive impact on the HTC prediction. This is because the increase in the temperature of the substrate promotes nucleation, thereby increasing the bubble frequency and favoring better heat removal from the surface \cite{MCHALE2010249, GOEL2017163}.

    \item Surface roughness ($R_q$) plays a major role in enhancing the nucleation site density. This increases the boiling performance \cite{KIM2016992}. The same variation can be observed from the SHAP interpretation analysis.

    \item Also, from the plot, the HTC increases when there is a decrease in the wettability of the surface. Since high wettability surfaces (low contact angle surfaces) increase the bubble departure radius and reduce the frequency of bubble emission, heat transfer decreases \cite{PHAN20095459}. On the other hand, low wettability surfaces (high contact angle surfaces) require low wall superheat to increase the bubble nucleation because of easier bubble detachment, hence promoting heat transfer \cite{MOZE2020120265}.

    \item When the effective area for heat transfer increases, heat transfer through the surface is also improved \cite{KUMAR2024125096}. The area augmentation factor indicates the increase in the effective area of microchannels when compared to a plain surface. Thus, surfaces having large $\lambda$ value augment the HTC.  

    \item The model predicts that HTC increases with an increase in fin height ($h_f$). As high-aspect-ratio ($h_f$/$w_f$) microchannels increase the capillary liquid-wicking behavior and increase the liquid flow to the dry spot, the boiling HTC is enhanced \cite{KWAK2018111}. 

    \item The increase in channel width ($w_g$) and fin width ($w_f$) have a negative impact on HTC enhancement, as observed in the SHAP analysis. Smaller width channels lead to a liquid jet impingement-like mechanism and facilitate liquid flow to the nucleation sites, thus improving heat transfer \cite{JAIKUMAR2016795, yao_2013}. Reducing the fin width enhances the heat transfer through the capillary wicking mechanism because of the increase in aspect ratio of the microchannel \cite{KWAK2018111, kalanifc8720212, kalaniethanol20212}. Also, reducing both $w_g$ and $w_f$ increases the area augmentation factor, resulting in a larger thermal energy transport \cite{kalanifc8720212, kalaniethanol20212}. The same phenomenon is observed for highly packed channels (i.e.) microchannels having a lesser pitch. For the same dimension, microchannels with smaller pitch have a large surface area for heat transfer, thereby enhancing HTC \cite{GHEITAGHY2017892}.

    \item The positive impact of liquid density on the model prediction can be attributed to the quick rewetting of the surface and enabling sustained heat transfer, particularly with liquid having a large density.

    \item $P_{film}$ represents the interfacial liquid pressure at the surface. When the pressure increases, the bubble departure diameter decreases, and the bubble departure frequency increases \cite{MAHMOUD2021101024}. This promotes heat transfer, and in this analysis, $P_{film}$ shows a positive influence on HTC prediction.

    \item Heat transfer through the liquid microlayer under the developing bubbles is enhanced when operating with liquids having higher thermal conductivity \cite{KUMAR2021100827}. Increased heat conduction through this layer enhances HTC, and thus, with an increase in $k_{l}$, the model has a direct impact on the model prediction.
  
\end{itemize}

\subsubsection{For the water and other fluids dataset}
The comparative analysis of the SHAP summary plot for water (Fig. \ref{fig: parallel_microchannels_bar_plot_water}) and other fluids (Fig. \ref{fig: parallel_microchannels_bar_plot_other_fluids}) dataset  summarizes the following:
\begin{itemize}
    \item Contact angle significance is more prominent in water than in fluids other than water. The high wetting tendency of the other fluids reduces the impact of the contact angle. However, surface roughness is one of the major contributing factors for both types of fluids. 

    \item Also, the surface characteristics - fin height, groove width, fin width, and area augmentation factor affect the model prediction significantly for both types of fluids.

    \item  The effect of wall superheat, or wall temperature, is more significant in fluids other than water. For datasets with water as the working fluid, the thermal conductivity of the liquid at the interface has a considerable effect.

    \item For water, in addition to the surface properties ($w_f$, $h_f$, $w_g$, $\lambda$), the thermal conductivity of the substrate ($k_w$) and pitch ($p$) of the channel displays their prominence. The substrate with high thermal conductivity boosts the HTC, as large thermal conductivity reduces the activation time for bubble nucleation, minimizes the reactivation time for bubble nucleation, and accelerates the bubble departure frequency \cite{AN2021107110}.

\end{itemize}

The above discussion illuminates the importance of each feature and its corresponding variation in the HTC prediction on microchannel structured surfaces. Also, most of the variations are validated against the existing experimental studies. Thus, SHAP interpretation provides transparency and corroborates the reliability of the proposed hybrid framework.

\section{Conclusions}
A model to predict the boiling heat transfer characteristics for microchannel structured surfaces is developed in this research. With 7128 data points amassed from available studies, an empirical correlation is proposed to predict the HTC. Then, a hybrid framework (ML + proposed correlation) is employed to improve the accuracy of HTC prediction. SHAP analysis has been inferred to identify the key parameters, and their variation impacting model prediction has been analyzed. 

The major conclusions include:

\begin{itemize}

    \item The available empirical correlations have been analyzed on the microchannel structured surfaces dataset. The Stephan-Preusser correlation exhibited a reasonable prediction. Then, a new correlation is proposed with the inclusion of the dimensionless parameters to the Stephan-Preusser correlation including $\lambda$, $\frac{k_w}{k_l}$, $\frac{R_q}{r_{\mathit{cav}}}$, $\frac{\theta}{90}$, $P_{\mathit{r}}$, $\frac{M_f}{M_w}$, $\frac{h_f}{w_f}$, $\frac{w_g}{\mathit{p}}$, and $\frac{D_h}{\mathit{p}}$. This correlation is able to show a better performance with 0.936 as the $R^2$ value than the original correlation with a $R^2$ value of only 0.55.
    
    \item Different ML models have been evaluated for HTC prediction on microchannel structured surfaces. ET, KNN, XGBoost, RF, and LightGBM models predicted HTC with a $R^2$ value of only around 0.91. DNN performance improved slightly with a $R^2$ value of 0.940. Then, a hybrid framework - PIMLAF is developed for the microchannel structured surfaces dataset, combining the proposed correlation and deep neural network model. The performance improved significantly with a $R^2$ value of 0.995.

    \item In the hybrid framework, the proposed correlation is able to provide a better starting point for the ML model. The integration of physics-based models and ML improves the model accuracy, generalizability, and robustness of the hybrid framework. The prior physics-informed prediction through the correlation enables the model to adhere to the primary principles of the boiling characteristics. This makes the hybrid model reliable and accurate in predicting unseen data.

    \item The parametric variation exhibited by the proposed model is analyzed through SHAP analysis and validated against the existing studies, thus trusting the model predictions even for inputs that fall beyond the trained parametric ranges. In this study, surface roughness ($R_q$), fin height ($h_f$), area augmentation factor ($\lambda$), and groove width ($w_g$) remain critical parameters affecting HTC across both types of fluids datasets and the overall dataset.

    \item Closely packed fins and high aspect ratio fins markedly improve the heat transfer. Closely packed configuration increases the area augmentation factor, and high-aspect ratio channels enhance the capillary wicking phenomena. Also, smaller width channels enhance $\lambda$ for a given dimension and promote liquid supply to the nucleation sites. Thus, the HTC is considerably improved with an increase in $h_f$ and $\lambda$ and with a decrease in $p$, $w_g$, and $w_f$.

    \item The contact angle plays a significant role in the water dataset. Its significance is not as significant in the other fluids as compared to water because the contact angle is almost the same for fluids other than water due to their large wetting tendency. However, the roughness of the surface is highly significant for both the fluid categories.
    
\end{itemize}

\bibliography{Microchannel_structured_surfaces_paper_arxiv.bib}
\bibliographystyle{elsarticle-num-names}


\addcontentsline{toc}{chapter}{References}
\clearpage


\end{document}